\documentclass[aps,prb,twocolumn,superscriptaddress,nofootinbib,longbibliography,eqsecnum]{revtex4-2}

\usepackage[T1]{fontenc}
\usepackage[utf8]{inputenc}
\usepackage{amsmath,amssymb,amsfonts,bm,braket,graphicx}
\usepackage{xcolor}
\usepackage{multirow}

\usepackage{amstext}
\usepackage{amsbsy}
\usepackage{url}
\usepackage{bm}
\usepackage{graphicx}
\usepackage{color}
\usepackage[colorlinks=true, urlcolor=blue, linkcolor=blue, citecolor=blue, pdftex]{hyperref}
\usepackage{mathrsfs}

\newcommand{\Jq}{\mathcal{J}_{\mathbf{q}}}
\newcommand{\Jzero}{\mathcal{J}_{\mathbf{0}}}
\newcommand{\dd}{\bm\delta}
\newcommand{\Ktot}{\mathbf{K}}
\newcommand{\qq}{\mathbf{q}}
\newcommand{\pp}{\mathbf{p}}
\newcommand{\rr}{\mathbf{r}}
\newcommand{\Aone}{\mathsf{A_{1}}}

\newcommand{\Btwo}{\mathsf{B_{2}}}

\newcommand{\Etwo}{\mathsf{E_{2}}}
\newcommand{\Etwopm}{\mathsf{E_{2}^\pm}}
\newcommand{\Etwop}{\mathsf{E_{2}^+}}
\newcommand{\Etwom}{\mathsf{E_{2}^-}}

\newcommand{\Econtmin}{E_{\text{cont}}^{\text{min}}}

\begin{document}

\title{Chiral enhancement of two-magnon bound states in an $S=1/2$ triangular-lattice magnet}

\author{László Rudner}
\affiliation{Department of Theoretical Physics, Institute of Physics, Budapest University of Technology and Economics,M\H{u}egyetem rakpart 3, H-1111 Budapest, Hungary}
\affiliation{HUN-REN Wigner Research Centre for Physics, Institute for Solid State Physics and Optics, Budapest, Hungary}

\author{Karlo Penc}
\affiliation{HUN-REN Wigner Research Centre for Physics, Institute for Solid State Physics and Optics, Budapest, Hungary}

\date{\today}

\begin{abstract}
We study one- and two-magnon excitations above the fully polarized state of the spin-\(1/2\) triangular-lattice \(J_1\)-\(J_2\)-\(J_3\)  Heisenberg model with a uniform scalar-chirality interaction. In the Heisenberg model, we identify two special manifolds of one-magnon minima by rewriting the dispersion in complete-square form. The chirality term cancels exactly in the one-magnon sector, leaving the dispersion and instability field unchanged, but survives in the two-magnon sector as an oriented interaction between neighboring magnons. Using symmetry-adapted lattice harmonics, we show that scalar chirality splits two-magnon states of opposite relative-motion chirality and selectively enhances the binding of one of them. It can strengthen an existing bound state, change the symmetry of the lowest magnon pair, or induce binding where none occurs without chirality. Exact diagonalization further shows that two-magnon pairing can also occur at finite total momentum. Our results identify scalar chirality as a microscopic mechanism for enhancing two-magnon pairing without modifying the one-magnon spectrum.
\end{abstract}

\maketitle

\section{Introduction}
\label{sec:introduction}

In a conventional magnetically ordered state, spin-rotational symmetry is broken by a dipolar order parameter, manifested as a finite expectation value of the spin operators themselves.
Quantum magnets can, however, support more subtle forms of symmetry breaking in which dipolar order is absent, and the order parameter is instead associated with higher-rank spin multipoles.
The simplest such example is spin-nematic order, characterized by a finite quadrupolar order parameter, which is even under time reversal.
For \(S=1\) moments, quadrupolar order can arise directly from local spin degrees of freedom, whereas in spin-\(1/2\) moments an on-site quadrupole is identically absent, and nematic order must instead reside in inter-site spin-bilinear operators
 \cite{1969nm, *1984nm, PencLauchli2011}. 
In this sense, spin-nematic order can be viewed as the two-particle counterpart of conventional magnon condensation: rather than a single-magnon mode becoming unstable, a bound state of two magnons softens and condenses while the one-magnon excitations remain gapped \cite{chubukov1991, Batista_RMP_2014}.
For \(S=1/2\), frustrated magnets with competing ferromagnetic nearest-neighbor and antiferromagnetic further-neighbor interactions provide natural settings for this mechanism.
Examples include the one-dimensional frustrated ferromagnetic chain, where quasi-long-range multipolar correlations emerge in a magnetic field \cite{kuzian2007, momoi1, Lauchli09, Zhitomirsky_2010, Leon_and_Oleg_2016}, and two-dimensional square-lattice models supporting genuine spin-nematic order \cite{shannon2006, Igbal_16, jiang2023where, gallegos_chernyshev}.
In these systems, frustration generates an effective attraction between magnons, allowing a two-magnon bound state to soften before the one-magnon mode becomes unstable and thereby driving a spin-nematic instability.

The high-magnetic-field regime provides a particularly clean setting for formulating this problem. Above the saturation field, the fully polarized state provides an exact vacuum for magnetic excitations, and the leading instability upon lowering the field is determined by which few-magnon excitation softens first.
If a single-magnon mode becomes unstable first, its condensation produces conventional transverse dipolar order.
If, instead, a two-magnon bound state softens before the one-magnon mode, its condensation signals a spin-nematic instability, with an order parameter of the form \(\langle S_i^- S_j^- \rangle\) for spin-\(1/2\) systems. 
This perspective also makes clear that the mere existence of a two-magnon bound state is not sufficient to establish a stable nematic phase.
One must determine whether the bound state actually preempts the one-magnon instability and whether the resulting phase remains stable against competing higher-multipolar or other many-body instabilities.
Recent work on square-lattice ferro-antiferromagnets has emphasized this point, showing that the parameter window for spin-nematic order can be strongly constrained once the full high-field phase diagram is taken into account \cite{jiang2023where,gallegos_chernyshev}.

Experimental evidence for spin-nematic phases remains scarce, despite the existence of several candidate materials \cite{Mila_17, exp2008sq, exp2011licuvo4, exp2017licuvo4, exp2017sq, exp2019sq, zheludev1, zheludev2, zheludev3, zheludev4, zheludev5}.
The main experimental challenge is that spin-nematic order is encoded in quadrupolar rather than dipolar spin correlations and therefore does not couple directly to conventional probes of single-spin fluctuations.
Nevertheless, two-magnon and quadrupolar excitations can leave characteristic signatures in inelastic neutron scattering \cite{Ribhu_2012,Wessel_2015,Smerald_Shannon_PRB_2013} and can be accessed more directly through multispin scattering vertices in Raman spectroscopy \cite{Shastry_Shraiman_PRL_1990,Michaud_2011}, including scalar-chirality channels on lattices with triangular loops \cite{Ko_PRB_2010}.
Resonant inelastic X-ray scattering (RIXS) provides another route to otherwise hidden multipolar correlations, including spin-nematic and scalar-chirality correlations \cite{savary2015probinghiddenordersresonant}.
Experiments on \(\mathrm{Y_2BaNiO_5}\), \(\mathrm{Sr_2IrO_4}\), and \(\mathrm{Sr_2CoGe_2O_7}\) illustrate how RIXS, Raman scattering, resonant diffraction, and electric-dipole-active spectroscopy can reveal quadrupolar correlations beyond conventional one-magnon probes \cite{Nag:2022aa,Kim:2024aa,Arima_2007,Mitsuru_2017}.

The triangular lattice provides a natural setting for revisiting the two-magnon problem. As the simplest two-dimensional non-bipartite Bravais lattice, it combines strong geometric frustration with elementary triangular plaquettes on which one can define the scalar spin chirality
\begin{equation}
  \hat \chi_{ijk} = \hat{\mathbf S}_i \cdot \left( \hat{\mathbf S}_j \times \hat{\mathbf S}_k \right),
  \label{eq:scalar_chirality}
\end{equation}
where \(\hat{\mathbf S}_i\) denotes the spin-\(1/2\) operator at lattice site \(i\) and the ordering \(i,j,k\) fixes the orientation of the plaquette.
The scalar chirality is invariant under global spin rotations, but changes sign under time reversal and under spatial symmetries that reverse the orientation of the triangle.
It plays an important role in theories of chiral spin liquids and related topological magnetic phases \cite{kalmeyer1987, *wen1989}. 
Chiral interactions and correlations on the triangular lattice have consequently been studied in a variety of extended spin and Hubbard-derived models. These include scalar-chirality interactions that stabilize Kalmeyer--Laughlin-type chiral spin liquids \cite{wietek2017}, triangular-lattice Hubbard models exhibiting chiral spin-liquid phases \cite{szasz2020, chen2022}, and multispin-exchange mechanisms proposed to generate such phases in Mott insulators \cite{cookmeyer2021}.

Beyond chiral physics, the triangular lattice also supports a broad range of multipolar magnetic orders.
For example, spin-\(1/2\) triangular-lattice models with multiple-spin exchange interactions, motivated in part by thin films of solid \(^3\)He, can support octupolar, or triatic, order \cite{momoi2006octupolar}, as well as octahedral spin-nematic order \cite{momoi2012spin}.
Spin-\(1\) models with biquadratic exchange likewise exhibit quadrupolar phases \cite{S1BBQ_PRL_2006, Tsunetsugu_2006, Subhro_2006}. 
On the materials side, the triangular antiferromagnet \(\mathrm{NiGa_2S_4}\) has long motivated spin-nematic scenarios \cite{NiGaS_Science_2005,Tsunetsugu_2006,Subhro_2006}, while recent experiments on the spin-\(1\) triangular-lattice compound \(\mathrm{Na_2BaNi(PO_4)_2}\) have reported two-magnon bound-state condensation and quadrupolar excitations associated with spin-nematic phases \cite{Sheng:2025aa,Na2BaNiPO42_PRL_2025}.
These results establish the triangular lattice as a fertile setting for multimagnon instabilities and multipolar magnetic order.
This raises a natural question: what role does scalar spin chirality play in the formation, symmetry, and stability of two-magnon bound states on the triangular lattice?

Here we address this question in the high-field limit of a spin-\(1/2\) triangular-lattice model with exchange interactions and a uniform scalar-chirality term.
We focus on the one- and two-magnon sectors above the fully polarized state, which provide a particularly transparent setting for isolating the effect of scalar chirality.
For this chirality pattern, we show that the scalar-chirality interaction leaves the one-magnon dispersion unchanged: in the fully polarized background, the hopping amplitudes generated by the two elementary triangles sharing each bond cancel exactly.
Consequently, neither the field at which the one-magnon gap closes nor the positions of the one-magnon minima are affected.
In the two-magnon sector, however, this cancellation is incomplete when the two flipped spins occupy neighboring sites.
The scalar-chirality interaction therefore contributes directly to the two-magnon interaction vertex, modifying the binding energy and lifting the degeneracy between states of opposite chirality.
In particular, it splits the \(\Etwo\)-type partial-wave sector into two components of opposite chirality and selectively enhances the binding of one of them.

The two-magnon problem represents the first level in a hierarchy of few-magnon instabilities near saturation.
A two-magnon bound state can drive a quadrupolar spin-nematic instability, but it may compete with bound states of three or more magnons associated with higher-rank multipolar order.
Such competition is particularly relevant on the spin-\(1/2\) triangular lattice, where multiple-spin-exchange models are already known to stabilize triatic and octahedral-nematic phases.
Establishing the ultimate high-field phase therefore requires extending the analysis to higher-magnon sectors.
The present work identifies the leading two-magnon bound-state channels and shows how scalar chirality reorganizes them by selectively favoring bound states of a definite chirality.

The remainder of this manuscript is organized as follows.
Section~\ref{sec:model} defines the spin-\(1/2\) triangular-lattice model, including the Heisenberg exchanges, the scalar-chirality interaction, and the lattice conventions.
In Sec.~\ref{sec:luttinger_tisza}, we revisit the Luttinger--Tisza phase diagram of the Heisenberg model and identify special one-magnon minimum manifolds by completing squares in the dispersion.
Section~\ref{sec:one_magnon} analyzes the one-magnon sector above the fully polarized state and shows how the scalar-chirality term cancels from the one-magnon dispersion.
Section~\ref{sec:two_magnon} derives the two-magnon Schr\"odinger equation, the scalar-chirality contribution to the two-magnon interaction kernel, and the corresponding gap equation for bound states.
The resulting bound-state spectra, binding energies, and dominant partial-wave channels are presented in Sec.~\ref{sec:results}.
Finally, Sec.~\ref{sec:conclusion} summarizes the main results and discusses their implications for high-field spin-nematic and multipolar instabilities.

\section{The spin Hamiltonian}
\label{sec:model}

\begin{figure}[tb]
    \centering
    \includegraphics[width=0.6\linewidth]{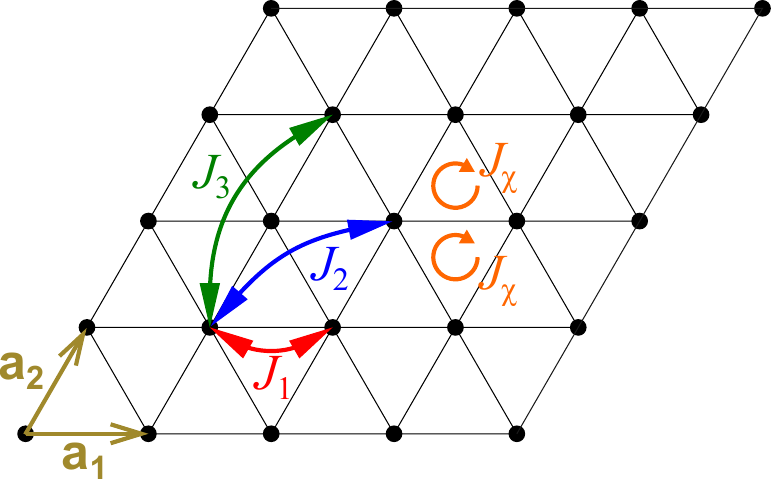}
    \caption{%
    Triangular lattice illustrating the \(J_1\) first-, \(J_2\) second-, and \(J_3\) third-neighbor Heisenberg exchange interactions, as well as the scalar-chirality terms \(\hat \chi_{ijk}\) defined on the up- and down-pointing triangles. The circular arrows indicate the orientation of the ordered sites \(i\), \(j\), and \(k\) around each triangle. The primitive lattice vectors are \(\mathbf a_1 = (1,0)\) and \(\mathbf a_2 = (1/2,\sqrt{3}/2)\).
    }
    \label{fig:lattice}
\end{figure}

We consider the spin-\(1/2\) Hamiltonian
\begin{equation}
   \hat H
   =
   \hat H_{\mathrm{Heis}}
   +
   \hat H_{\mathrm{Chir}}
   +
   \hat H_{\mathrm{Zeem}} .
   \label{eq:hamiltonian}
\end{equation}

The Heisenberg part is
\begin{multline}
    \hat H_{\mathrm{Heis}}
    =
    \frac{J_1}{2}
    \sum_{\rr}
    \sum_{\mu=1}^{6}
    \hat{\mathbf S}_{\rr}
    \cdot
    \hat{\mathbf S}_{\rr+\dd_\mu^{(1)}}
    \\
    +
    \frac{J_2}{2}
    \sum_{\rr}
    \sum_{\mu=1}^{6}
    \hat{\mathbf S}_{\rr}
    \cdot
    \hat{\mathbf S}_{\rr+\dd_\mu^{(2)}}
    \\
    +
    \frac{J_3}{2}
    \sum_{\rr}
    \sum_{\mu=1}^{6}
    \hat{\mathbf S}_{\rr}
    \cdot
    \hat{\mathbf S}_{\rr+\dd_\mu^{(3)}} .
    \label{eq:hamiltonian_Heis}
\end{multline}
Here \(\rr\) runs over all lattice sites. The vectors \(\dd_\mu^{(n)}\) connect a site to its \(n\)th-neighbor sites, as defined in Eqs.~\eqref{eq:dd1}--\eqref{eq:dd3}. Each shell contains six bond vectors, and the factor \(1/2\) removes the double counting of bonds. For \(S=1/2\), the exchange operator can be written in terms of the two-site permutation operator \(\hat P_{ij}\) as
\begin{equation}
  \hat{\mathbf S}_i \cdot \hat{\mathbf S}_j
  =
  \frac{1}{2}\hat P_{ij}
  -
  \frac{1}{4}.
\end{equation}

The scalar-chirality term is
\begin{equation}
    \hat H_{\mathrm{Chir}}
    =
    4 J_{\chi}
    \sum_{\langle ijk\rangle_{\bigtriangleup,\bigtriangledown}}
    \hat\chi_{ijk},
    \label{eq:Ham_chirality}
\end{equation}
where \(\hat\chi_{ijk}\) is defined in
Eq.~\eqref{eq:scalar_chirality}.
The sum runs once over all elementary up- and down-pointing triangles,
with the same cyclic orientation assigned to every triangle, as
indicated by the circular arrows in Fig.~\ref{fig:lattice}.
This convention defines the uniform chirality pattern considered
throughout.
Reversing all triangle orientations is equivalent to changing
\(J_\chi\to-J_\chi\).
The factor of \(4\) in Eq.~\eqref{eq:Ham_chirality} is chosen such that,
for spin \(1/2\), \(J_\chi\) is the coefficient of the antisymmetric
cyclic-exchange operator. Indeed,
\begin{equation}
    \hat\chi_{ijk}
    =
    \frac{i}{4}
    \left(
      \hat P_{ijk}
      -
      \hat P_{ikj}
    \right),
    \label{eq:chirality_permutation}
\end{equation}
where \(\hat P_{ijk}\) cyclically permutes the spins as
\(i\to j\to k\to i\), while
\(\hat P_{ikj}=\hat P_{ijk}^{-1}\) denotes the opposite cyclic
orientation.
Equation~\eqref{eq:Ham_chirality} can therefore equivalently be written as
\begin{equation}
    \hat H_{\mathrm{Chir}}
    =
    i J_{\chi}
    \sum_{\langle ijk\rangle_{\bigtriangleup,\bigtriangledown}}
    \left(
      \hat P_{ijk}
      -
      \hat P_{ikj}
    \right).
\end{equation}

Such a term arises at third order in the strong-coupling expansion of a
Hubbard model when the electron hopping amplitudes acquire a nonzero
gauge-invariant Peierls phase around a triangular plaquette, corresponding
to a finite flux through the plaquette
\cite{Motrunich2006,Bulaevskii2008}.
The scalar chirality is odd under time reversal and under spatial
operations that reverse the orientation of a triangle.
For fixed \(J_\chi\neq0\), the chirality interaction therefore breaks
time-reversal and orientation-reversing lattice symmetries while
preserving the proper rotations.
In terms of planar point groups, the spatial symmetry is reduced from
\(D_6\) to \(C_6\).

Finally, the Zeeman term is
\begin{equation}
   \hat H_{\mathrm{Zeem}}
   =
   -h
   \sum_{\rr}
   \hat S^z_{\rr},
   \label{eq:Ham_Zeem}
\end{equation}
corresponding to an external magnetic field along the \(z\) axis. Both \(\hat H_{\mathrm{Heis}}\) and \(\hat H_{\mathrm{Chir}}\) are \(\mathrm{SU}(2)\) symmetric, whereas \(\hat H_{\mathrm{Zeem}}\) reduces the spin-rotation symmetry to \(\mathrm{U}(1)\).

\section{Luttinger--Tisza analysis of the Heisenberg model}
\label{sec:luttinger_tisza}

We first consider the Heisenberg part of Eq.~\eqref{eq:hamiltonian}.
The triangular-lattice \(J_1\)-\(J_2\)-\(J_3\) Heisenberg model supports a variety of commensurate and incommensurate magnetic states for both antiferromagnetic and ferromagnetic nearest-neighbor exchange \cite{Messio_2011,Okubo_PRL_2012,Leonov:2015aa}. Recent Luttinger--Tisza and numerical studies have emphasized, in particular, the degenerate spiral manifolds that occur in the frustrated antiferromagnetic regime for  \(J_2=2J_3\)  \cite{Mohylna_2022}.

\begin{figure}[tb]
    \centering
    \includegraphics[width=0.6\linewidth]{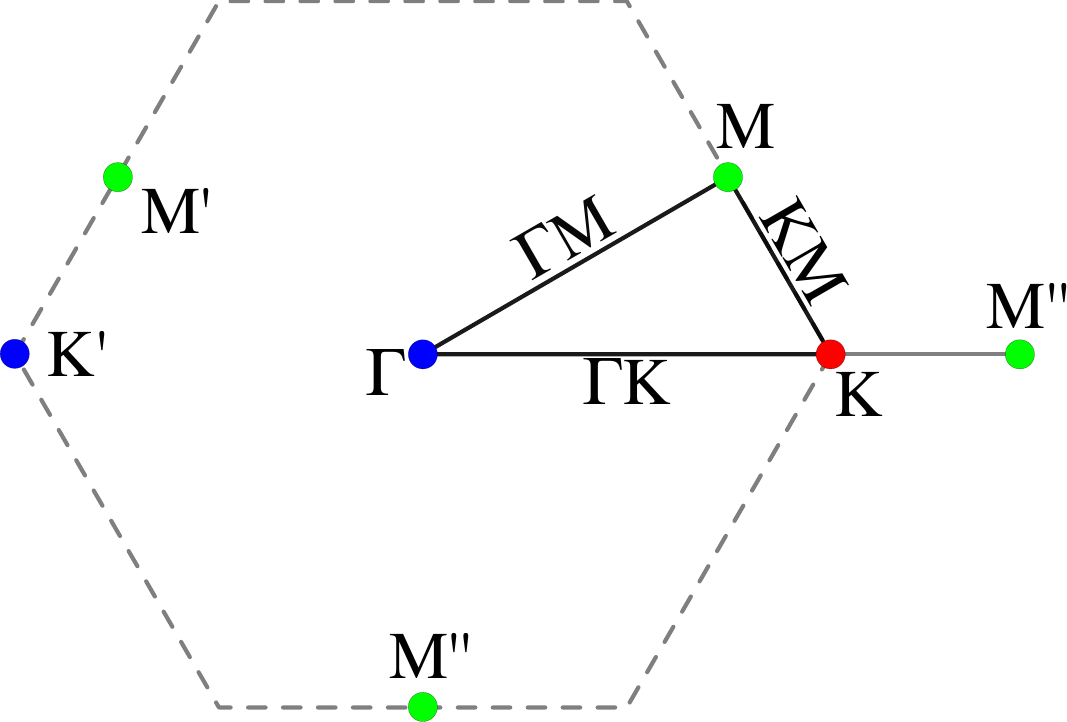}
    \caption{%
    Brillouin zone of the triangular lattice showing the inequivalent \(\Gamma\), \(K\), and \(M\) points, together with high-symmetry lines. The \(\qq = (q,0)\) line defined in Eq.~\eqref{eq:GKMline} is also indicated; it connects \(\Gamma\), \(K\), and \(M''\) points along a straight path.
    }
    \label{fig:bz}
\end{figure}

For a Bravais lattice with isotropic exchange, the Luttinger--Tisza method \cite{TiszaLuttinger} reduces the classical ordering problem to minimizing
the Fourier-transform of the exchange interaction 
\begin{equation}
    \Jq=6J_{1}\gamma^{(1)}_{\qq}+6J_{2}\gamma^{(2)}_{\qq}+6J_{3}\gamma^{(3)}_{\qq}.
    \label{eq:J_gamma}
\end{equation}
 over the Brillouin zone. 
The \(\gamma^{(n)}_{\qq}\) above are the lattice harmonics defined in Eq.~\eqref{eq:lattice_harmonics_A1}. Explicitly, \(\Jq\) takes the form
\begin{align}
\Jq={}&2J_{1}\left[\cos q_{x}+2\cos\frac{q_{x}}{2}\cos\frac{\sqrt{3}q_{y}}{2}\right]
\nonumber\\
&+2J_{2}\left[\cos(\sqrt{3}q_{y})+2\cos\frac{3q_{x}}{2}\cos\frac{\sqrt{3}q_{y}}{2}\right]
\nonumber\\
&+2J_{3}\left[\cos(2q_{x})+2\cos q_{x}\cos(\sqrt{3}q_{y})\right].
\label{eq:Jq}
\end{align}
At the high-symmetry points in the Brillouin zone, Fig.~\ref{fig:bz},
\begin{equation}
\Gamma=(0,0),\quad K=\left(\frac{4\pi}{3},0\right),\quad
M=\left(\pi,\frac{\pi}{\sqrt{3}}\right)
\end{equation}
the \(\Jq\) takes the following values:
\begin{subequations}
\begin{align}
    \mathcal{J}_{\Gamma}
    &= 6 J_1+ 6 J_2+ 6J_3,
    \label{eq:J_Gamma}
    \\
    \mathcal{J}_{K}
    &= -3J_1+6J_2-3J_3,
    \label{eq:J_K}
    \\
    \mathcal{J}_{M}
    &= -2J_1-2J_2+6J_3.
    \label{eq:J_M}
\end{align}
\end{subequations}

\subsection{Stability of the ferromagnetic phase}

The ferromagnetic state corresponds to a minimum at \(\Gamma\).
The energy of the ferromagnetic \(\Gamma\)-point state becomes degenerate with that of the commensurate (K)-point state when \(\mathcal{J}_K=\mathcal{J}_\Gamma\), which gives
\begin{equation}
J_1+J_3=0.
\label{eq:FM_K_boundary}
\end{equation}
Similarly, degeneracy between the \(\Gamma\)- and \(M\)-point states occurs when \(\mathcal{J}_M=\mathcal{J}_\Gamma\), yielding
\begin{equation}
J_1+J_2=0.
\label{eq:FM_M_boundary}
\end{equation}
On these two boundaries, the additional soft modes are located at the full symmetry stars of \(K\) and \(M
\), respectively.

There is also a long-wavelength instability of the ferromagnetic state.
Expanding around the \(\Gamma\) point in the Brillouin zone, one finds
\begin{align}
    \Jq-\mathcal{J}_{\Gamma}
    ={}&
    -\frac{3}{2}
    \left(
        J_1+3J_2+4J_3
    \right)
    |\bm q|^2
    \nonumber\\
    &+
    \frac{3}{32}
    \left(
        J_1+9J_2+16J_3
    \right)
    |\bm q|^4
    \nonumber\\
    &-
    \frac{1}{384}
    \left(
        J_1+27J_2+64J_3
    \right)
    |\bm q|^6
    \nonumber\\
    &-
    \frac{1}{3840}
    \left(
        J_1-27J_2+64J_3
    \right)
    |\bm q|^6
    \cos\left(6\theta_{\bm q}\right)
    +
    O(q^8),
    \label{eq:Jq_Gamma_expansion}
\end{align}
where \(\theta_{\bm q}\) denotes the polar angle of \(\bm q\).
Thus, the spin stiffness vanishes on the Lifshitz line
\begin{equation}
    J_1+3J_2+4J_3=0.
    \label{eq:FM_Lifshitz_boundary}
\end{equation}
On this line, the leading dispersion is generically quartic. Using
\(J_1=-3J_2-4J_3\), its coefficient becomes
\begin{equation}
    \frac{3}{32}
    \left(
        J_1+9J_2+16J_3
    \right)
    =
    \frac{9}{16}
    \left(
        J_2+2J_3
    \right).
\end{equation}
For \(J_1=-1\), the quartic coefficient also vanishes at
\((J_2,J_3)=(1,-1/2)\), so that the leading dispersion near \(\Gamma\)
is of sixth order.  As discussed below in
Sec.~\ref{sec:codimension-one}, this point exhibits an extended
nodal-line manifold of minima.

Combining these conditions, the full Luttinger--Tisza stability region
of the ferromagnetic state in the \(J_1\)-\(J_2\)-\(J_3\) triangular-lattice
model is
\begin{subequations}
\label{eq:FM_LT_region}
\begin{align}
    J_1+3J_2+4J_3 &\le 0, \\
    J_1+J_2 &\le 0, \\
    J_1+J_3 &\le 0.
\end{align}
\end{subequations}
%

\subsection{Commensurate and spiral phases}
\label{sec:ordered_phases}

Away from the ferromagnetic region, the ordering wave vector is obtained by comparing the local minima of \(\Jq\).
The minimum at the commensurate K-point realizes the usual three-sublattice order, while the minimum at the M point is either a stripy phase (single-\(\qq\)) or a four-sublattice order (triple-\(\qq\) structure). 

But the minima can also occur along the high-symmetry lines.
On the \(\Gamma M\) line, parametrized by
\begin{equation}
    \bm q=\left(q,\frac{q}{\sqrt3}\right),
    \qquad 0\le q\le \pi,
\end{equation}
the non-endpoint extremal point satisfies
\begin{equation}
    \cos q_{\Gamma M}
    =
    -\frac{J_1+J_2}{2J_2+4J_3},
    \label{eq:qGM}
\end{equation}
provided \(J_2+2J_3\neq0\) and the right-hand side lies in
\([-1,1]\).
This corresponds to 
\begin{equation}
 \mathcal{J}_{\Gamma M}=6 J_1 + 6 J_2 + 6 J_3 - \frac{(J_1 + 3 J_2 + 4 J_3)^2}{J_2 + 2 J_3}.
\end{equation}

\begin{figure}[bt]
    \centering
    \includegraphics[width=0.68\linewidth]{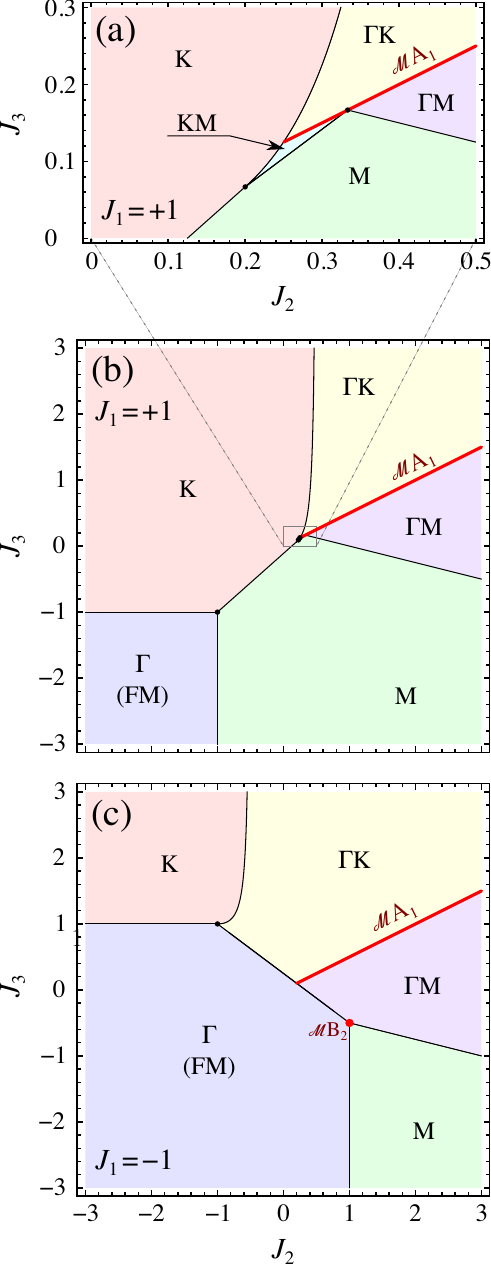}
\caption{%
Luttinger--Tisza phase diagram of the triangular-lattice \(J_1\)-\(J_2\)-\(J_3\) Heisenberg model.
Panels (a) and (b) correspond to the antiferromagnetic nearest-neighbor case, \(J_1 = +1\), with panel (a) giving an enlarged view of the central region of panel (b). Panel (c) shows the ferromagnetic nearest-neighbor case, \(J_1 = -1\).
Colors indicate the ordering wave vector \(\qq\) that minimizes \(\Jq\). They distinguish the commensurate \(\Gamma\), \(K\), and \(M\) regions from incommensurate spiral regions whose minima lie on the high-symmetry lines \(\Gamma K\), \(\Gamma M\), and \(KM\).
Equivalently, these wave vectors are the momenta at which the one-magnon dispersion softens.
The thick red line marks the \(J_2 = 2J_3\) locus associated with the \(\mathcal{M}_{\Aone}\) minimum manifold [Eq.~\eqref{eq:MA1_def}], while the red circle in panel (c) marks the fine-tuned point \((J_1,J_2,J_3) = (-1,1,-1/2)\), where the \(\mathcal{M}_{\Btwo}\) nodal-line minimum manifold is realized [Eq.~\eqref{eq:MB2_def}].
}
  \label{fig:Luttinger_Tisza_PD}
\end{figure}

It is useful to treat the \(\Gamma K\) line and the adjacent \(KM\) edge in an extended-zone scheme, as illustrated in Fig.~\ref{fig:bz}. 
We parametrize
\begin{equation}
 \qq=(q,0),\qquad 0\le q\le 2\pi .
 \label{eq:GKMline}
\end{equation}
This path starts at \(\Gamma\), passes through \(K\) at \(q=4\pi/3\), and ends at a symmetry-equivalent \(M\) point at \(q=2\pi\).
The condition for the minimum (extremum) factorizes as
\begin{multline}
\frac{d\mathcal{J}_{(q,0)}}{dq}
=
- 2 \sin \frac{q}{2}
  \left(1 + 2 \cos\frac{q}{2}\right) \\
  \times\left[J_1- 3 J_2 + (6 J_2 - 4 J_3) \cos \frac{q}{2} + 
   8 J_3 \cos^2 \frac{q}{2} \right].
 \label{eq:GK_extended_stationarity}
\end{multline}
The first factor gives the endpoints \(q=0\) and \(q=2\pi\), corresponding to \(\Gamma\) and \(M\), respectively. The second factor gives the \(K\) point, \(\cos (q/2)=-1/2\), i.e. \(q=4\pi/3\). The remaining minimal points are determined by
\begin{equation}
J_1- 3 J_2 + (6 J_2 - 4 J_3) \cos \frac{q}{2} + 
   8 J_3 \cos^2 \frac{q}{2} = 0.
\label{eq:GK_extended_quadratic}
\end{equation}
Thus, for \(J_3\ne0\),
\begin{equation}
\cos \frac{q_{\pm}}{2}
=
\frac{
2J_3-3J_2
\pm
\sqrt{(3J_2+2J_3)^2-8J_1J_3}
}
{8J_3}.
\label{eq:qGK_extended_x}
\end{equation}
The Luttinger--Tisza phase diagram is then obtained by keeping only those stationary points that are true minima and comparing their values of \(\Jq\).
In Fig.~\ref{fig:Luttinger_Tisza_PD}, we show the ordering vectors separately for ferromagnetic (\(J_1=-1\)) and antiferromagnetic (\(J_1=1\)) nearest-neighbor exchange.

\subsection{Codimension-one minimum manifolds}
\label{sec:codimension-one}

We now identify two cases in which the minima of \(\Jq\) form
codimension-one manifolds in momentum space.

\subsubsection{\(\Aone\) manifold}

An exact degeneracy occurs on the line \(J_2=2J_3\) as a consequence
of the identity between the triangular-lattice harmonics
\begin{equation}
    2\gamma^{(2)}_{\bf q}
    +
    \gamma^{(3)}_{\bf q}
    =
    6\left[\gamma^{(1)}_{\bf q}\right]^2
    -
    2\gamma^{(1)}_{\bf q}
    -
    1.
    \label{eq:gamma_identity}
\end{equation}
Therefore, on \(J_2=2J_3\), the exchange Fourier transform can be
written in the complete-square form
\begin{equation}
    \Jq
    =
    18J_2
    \left[
        \gamma^{(1)}_{\bf q}
        +
        \frac{J_1-J_2}{6J_2}
    \right]^2
    -
    \frac{(J_1-J_2)^2}{2J_2}
    -
    3J_2.
    \label{eq:J_square_J2_2J3_gamma}
\end{equation}
For \(J_2>0\), the coefficient of the square is positive. Provided the
value required to make the square vanish lies within the range of the
nearest-neighbor harmonic, the global minima form the contour
\begin{equation}
    \mathcal{M}_{\Aone}
    =
    \left\{
        \qq\in{\rm BZ}
        \,\middle|\,
        \gamma^{(1)}_{\qq}
        =
        \frac{J_2-J_1}{6J_2}
    \right\}.
    \label{eq:MA1_def}
\end{equation}
Here the subscript indicates that the manifold is defined as a level set
of the fully symmetric \(\Aone\) harmonic \(\gamma^{(1)}_{\qq}\).
Since
\begin{equation}
    -\frac{1}{2}
    \le
    \gamma^{(1)}_{\qq}
    \le
    1,
\end{equation}
the contour exists when
\begin{equation}
    -\frac{1}{2}
    \le
    \frac{J_2-J_1}{6J_2}
    \le
    1.
    \label{eq:degenerate_contour_condition}
\end{equation}
For \(J_1>0\), the allowed portion of this line corresponds to the spiral-spin-liquid manifold discussed in 
Ref.~\cite{Mohylna_2022}.
For \(J_1<0\), the same square completion gives the corresponding
ferromagnetic-side spiral contour once the \(\Gamma\) minimum becomes
unstable.
For example, for \(J_1=-1\) and \(J_2=2J_3>0\), the contour emerges
from \(\Gamma\) at \(J_2=1/5\), where the condition in
Eq.~\eqref{eq:degenerate_contour_condition} is first satisfied.
This coincides with the Lifshitz boundary
Eq.~\eqref{eq:FM_Lifshitz_boundary}; for \(J_2>1/5\), the degenerate
minimum manifold is a finite contour surrounding \(\Gamma\).

The complete-square form in Eq.~\eqref{eq:J_square_J2_2J3_gamma}
shows that the line \(J_2=2J_3\) provides a triangular-lattice
realization of the general codimension-one spiral-manifold construction
for Bravais-lattice Heisenberg models~\cite{Balla_PRB_2019}.
In two dimensions, the Luttinger--Tisza minima are determined by a
single scalar constraint on \(\qq\) and therefore form a one-dimensional
contour in the Brillouin zone.

\subsubsection{\(\Btwo\)  manifold}

A second exact codimension-one minimum manifold occurs at the fine-tuned point
\((J_1,J_2,J_3)=(-1,1,-1/2)\). This degeneracy is naturally expressed in terms of a one-dimensional nontrivial lattice harmonic. With the bond convention of Appendix~\ref{sec:trilatt}, define
\begin{equation}
\beta^{(1)}_{\qq}
=
\frac{1}{3}
\left[
\sin(\qq\cdot\boldsymbol\delta_1^{(1)})
-
\sin(\qq\cdot\boldsymbol\delta_2^{(1)})
+
\sin(\qq\cdot\boldsymbol\delta_3^{(1)})
\right].
\label{eq:B2_harmonic_def}
\end{equation}
Equivalently,
\begin{equation}
\beta^{(1)}_{\qq}
=
\frac{2}{3}
\sin\frac{q_x}{2}
\left[
\cos\frac{q_x}{2}
-
\cos\frac{\sqrt3 q_y}{2}
\right].
\label{eq:B2_harmonic_explicit}
\end{equation}
With the mirror convention adopted in Appendix~\ref{sec:trilatt},
the harmonic \(\beta^{(1)}_{\qq}\) transforms as \(\Btwo\).

 At the special point, the exchange Fourier transform takes the exact square form
\begin{equation}
\Jq
=
-3
+
18\left[\beta^{(1)}_{\qq}\right]^2 .
\label{eq:J_square_B2_point}
\end{equation}
Hence the Luttinger--Tisza minima are given by the nodal manifold
\begin{equation}
\mathcal{M}_{\Btwo}
=
\left\{
\qq\in{\rm BZ}
\,\middle|\,
\beta^{(1)}_{\qq}=0
\right\},
\label{eq:MB2_def}
\end{equation}
or explicitly,
\begin{equation}
\sin\frac{q_x}{2} = 0
\quad \mathrm{or} \quad
\cos\frac{q_x}{2} - \cos\frac{\sqrt3 q_y}{2} = 0.
\label{eq:B2_nodal_condition}
\end{equation}
Thus, \(\mathcal{M}_{\Btwo}\) is a union of symmetry-related lines in the
Brillouin zone.
This exact nodal-line structure also explains the anomalously soft
dispersion near \(\Gamma\) found above. Expanding
Eq.~\eqref{eq:J_square_B2_point} around \(\Gamma\) gives
\begin{equation}
    \Jq-\mathcal{J}_{\Gamma}
    =
    \frac{1}{64}
    |\bm q|^6
    \left[
        1+\cos\left(6\theta_{\bm q}\right)
    \right]
    +
    O(q^8),
\end{equation}
where \(\theta_{\bm q}\) denotes the polar angle of \(\bm q\).
The absence of quadratic and quartic terms is therefore the local
manifestation near \(\Gamma\) of the \(\mathcal{M}_{\Btwo}\) nodal-line
minimum manifold: the sixth-order term vanishes along the nodal
directions passing through \(\Gamma\).
This fine-tuned point thus provides a second triangular-lattice
realization of a codimension-one minimum manifold in the sense of
Ref.~\cite{Balla_PRB_2019}, distinct from the \(J_2=2J_3\) case, where
the square completion involves the fully symmetric \(\Aone\) harmonic
\(\gamma^{(1)}_{\qq}\).

In Fig.~\ref{fig:Luttinger_Tisza_PD}, the parameter-space loci associated
with the two minimum manifolds are indicated. The portion of the \(J_2=2J_3\) line
supporting \(\mathcal{M}_{\Aone}\) is shown in red, while the fine-tuned
point \((J_2,J_3)=(1,-1/2)\), where \(\mathcal{M}_{\Btwo}\) occurs for
\(J_1=-1\), is marked by a red circle.

\section{One-magnon sector}
\label{sec:one_magnon}

Here we revisit the same exchange Fourier transform from the viewpoint needed for the high-field problem: we identify the ferromagnetic stability boundaries and relate the Luttinger-Tisza minima directly to the one-magnon dispersion above the fully polarized state.

\subsection{Heisenberg interactions}

Let us denote the fully polarized state by \(|{\rm FM}\rangle\). For the Heisenberg model \eqref{eq:hamiltonian_Heis}, its energy is 
\begin{equation}
    E_0
    =
    \langle {\rm FM}|\hat H_{\mathrm{Heis}}|{\rm FM}\rangle
    =
    \frac12 NS^2\Jzero.
    \label{eq:E0}
\end{equation}
A normalized one-magnon Bloch state is
\begin{equation}
    |\psi^{(1)}_{\bm q}\rangle
    =
    \frac{1}{\sqrt{2SN}}
    \sum_j
    e^{-i\qq\cdot\mathbf{r}_j}
    \hat S_j^-|{\rm FM}\rangle.
    \label{eq:one_magnon_state}
\end{equation}
Direct evaluation gives
\begin{equation}
    \varepsilon_{\bm q}
    =
    E_{\bm q}-E_0
    =
    S\left(
    \Jq
    -
    \Jzero
    \right).
    \label{eq:one_magnon_dispersion}
\end{equation}
Thus, the one-magnon soft modes are precisely the Luttinger--Tisza minima of \(\Jq\), measured relative to the ferromagnetic point \(\Gamma\).

At the \(K\) and \(M\) points, the one-magnon energies are
\begin{align}
    \varepsilon_K
    &=
    -9S(J_1+J_3),
    \label{eq:eps_K}
    \\
    \varepsilon_M
    &=
    -8S(J_1+J_2),
    \label{eq:eps_M}
\end{align}
whereas near \(\Gamma\),
\begin{equation}
    \varepsilon_{\bm q}
    =
    -\frac{3S}{2}
    \left(
        J_1+3J_2+4J_3
    \right)
    |\bm q|^2
    +
    O(q^4),
    \label{eq:eps_Gamma_expansion}
\end{equation}
which is the one-magnon counterpart of
Eq.~\eqref{eq:Jq_Gamma_expansion}.
The stability of the fully polarized state against one-magnon
excitations at zero field is therefore equivalent to the
Luttinger--Tisza stability condition in Eq.~\eqref{eq:FM_LT_region}.

In a magnetic field, a one-magnon excitation acquires an additional
Zeeman energy \(h\), and the dispersion becomes
\begin{equation}
    \varepsilon_{\bm q}(h)
    =
    h
    +
    S\left(
        \Jq-\Jzero
    \right).
    \label{eq:one_magnon_field}
\end{equation}
It follows that the one-magnon instability field is
\begin{equation}
    h_c^{(1)}
    =
    S\left[
        \Jzero
        -
        \min_{\bm q}\Jq
    \right].
    \label{eq:critical_field_one_magnon}
\end{equation}
If the leading instability of the saturated state occurs in the one-magnon channel, magnons condense at the minimizing wave vector when the field is lowered through \(h_c^{(1)}\), producing transverse dipolar order \cite{Batista_RMP_2014}.
The two-magnon calculation below determines whether a bound state becomes unstable at a higher field instead.

\subsection{Scalar spin chirality}
\label{sec:chi_one_magnon}

\begin{figure}[tb]
    \centering
    \includegraphics[width=0.9\linewidth]{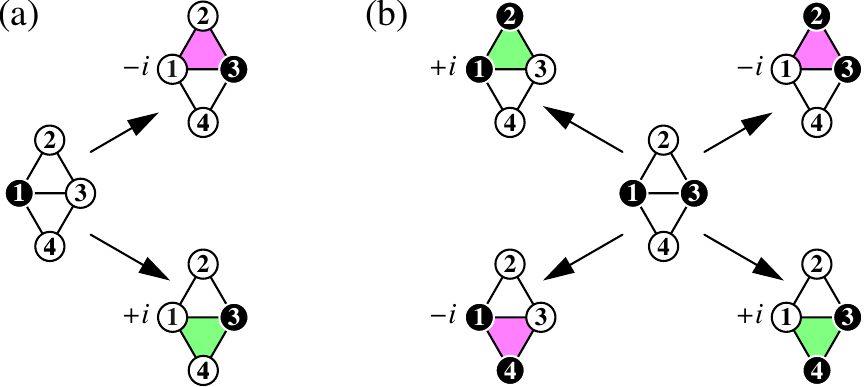}
\caption{%
Action of the uniform scalar spin chirality interaction \(4( \hat\chi_{123}+ \hat\chi_{134}) = i \hat P_{123} - i \hat P_{132} + i \hat P_{134} - i \hat P_{143}\) for spin-\(1/2\) one- and two-magnon states; see Eq.~\eqref{eq:chirality_permutation}.
Empty circles denote up spins and filled circles down spins. Colored triangles indicate the permutation terms that act nontrivially on the displayed spin configuration; the color encodes the sign of the imaginary amplitude, with magenta denoting \(-i\) and green \(+i\).
(a) In the one-magnon sector, the down spin on site \(1\) can hop to site \(3\) through two oriented triangles, corresponding to the processes \(-i\hat P_{132}\) and \(+i\hat P_{134}\). These two amplitudes cancel exactly, so the scalar-chirality term does not contribute to the one-magnon dispersion.
(b) In the two-magnon sector, a neighboring down spin prevents this cancellation, resulting in uncompensated imaginary amplitudes. These amplitudes originate from the time-reversal-odd scalar-chirality operator and act as a chiral interaction between magnons.
}
\label{fig:chiral_hoppings}
\end{figure}

Unlike the bilinear Heisenberg interactions, the scalar-chirality term cannot be represented by a Fourier-transformed exchange \(\mathcal J_{\qq}\) and therefore does not enter the Luttinger--Tisza analysis above.
We instead evaluate its action directly in the few-magnon sectors above the fully polarized state.

For \(S=1/2\), Eq.~\eqref{eq:chirality_permutation} expresses the scalar chirality on an oriented triangle in terms of cyclic permutations. Since both cyclic permutations act identically on the fully polarized state, the scalar-chirality contribution to its energy vanishes exactly. This representation also makes the cancellation in the one-magnon sector transparent. 
For the uniform chirality pattern considered here, a single flipped spin
can hop through the two oriented triangles sharing a bond, but the two
amplitudes have opposite signs.
As illustrated in Fig.~\ref{fig:chiral_hoppings}, the processes
\(-i\hat P_{132}\) and \(+i\hat P_{134}\) both move the down spin from
site \(1\) to site \(3\), and their amplitudes cancel exactly.
Thus, the scalar-chirality term annihilates both the fully polarized
state and the entire one-magnon subspace,
\begin{equation}
    \hat H_{\mathrm{Chir}}
    |\psi^{(1)}_{\qq}\rangle
    =
    0.
\end{equation}
Consequently, it leaves the one-magnon dispersion Eq.~\eqref{eq:one_magnon_dispersion}, the one-magnon
instability field \(h_c^{(1)}\), and the soft-mode wave vectors
unchanged. This statement is restricted to the magnon spectrum above the fully polarized state, and does not amount to a Luttinger--Tisza treatment of the full Hamiltonian. 

The two-magnon sector is different.
When two neighboring spins are flipped, the local pairwise cancellation
present for an isolated magnon is incomplete.
The scalar-chirality term then produces uncompensated oriented imaginary
amplitudes, which enter the two-magnon problem as an interaction vertex.
These oriented imaginary amplitudes distinguish opposite senses of
relative motion of the two magnons, providing the microscopic origin
of the chiral splitting derived below.
We derive its explicit contribution to the two-magnon Schr\"odinger
equation below.

This mechanism is related to the kinetic origin of magnon pairing in frustrated ferromagnets with cyclic exchange, where neighboring flipped spins can propagate coherently even when single-magnon motion is suppressed~\cite{shannon2006}.
Here the pair dynamics is instead generated by the time-reversal-odd combination \(i\hat P-i\hat P^{-1}\), providing a chiral analog of
magnon-pair hopping.

\section{Two-magnon sector}
\label{sec:two_magnon}

We now turn to the two-magnon problem above the fully polarized state, restricting throughout to spin-\(1/2\). This is the minimal sector in which magnon interactions appear. The problem of two-spin-wave bound states in ferromagnets has a long history~\cite{Wortis_63,Hanus_1963,Mattis}. In frustrated ferromagnets, such bound states can soften before the one-magnon mode and drive multipolar, in particular spin-nematic, instabilities~\cite{chubukov1991,shannon2006,momoi4,momoi2012spin,Zhitomirsky_2010,jiang2023where,gallegos_chernyshev}.

Our derivation follows Ref.~\cite{Zhitomirsky_2010,jiang2023where}, adapted here to the \(S=1/2\) triangular-lattice \(J_1\)-\(J_2\)-\(J_3\) model, including the scalar-chirality interaction. We first write the two-magnon Schrödinger equation in momentum space, where the one-magnon hopping determines the kinetic part and the interaction enters through a two-particle scattering kernel. We then express this kernel in a finite basis of lattice harmonics, or partial-wave channels, appropriate to the finite range of the interactions. This partial-wave representation reduces the two-magnon bound-state problem to a finite-dimensional eigenvalue equation.

\subsection{Two-magnon Schrödinger equation for the Heisenberg terms}
\label{sec:two-magnon-Schrodinger}

A configuration-space basis state in the \(S=1/2\) two-magnon Hilbert
space above the fully polarized state is
\begin{equation}
    \ket{i,j}
    =
    \hat S_j^- \hat S_i^- \ket{\mathrm{FM}},
    \qquad
    i\neq j,
    \label{eq:ij}
\end{equation}
where \(i\neq j\) enforces the hard-core constraint.
We introduce the non-normalized two-magnon momentum states
\begin{equation}
    \ket{\mathbf k,\mathbf k'}_{\mathrm R}
    =
    \sum_{i\neq j}
    e^{-i\mathbf k\cdot\mathbf r_i}
    e^{-i\mathbf k'\cdot\mathbf r_j}
    \ket{i,j} ,
    \label{eq:two_magnon_state}
\end{equation}
where the subscript \({\mathrm R}\) stands for the `regular' states respecting the hard-core constraint. In terms of the total momentum
\begin{equation}
    \Ktot = \mathbf k+\mathbf k'
\end{equation}
and relative momentum
\begin{equation}
    \qq = \frac{\mathbf k'-\mathbf k}{2},
\end{equation}
this state becomes
\begin{equation}
    \ket{\Ktot,\qq}_{\mathrm R}
    =
    \sum_{i\neq j}
    e^{-i\frac{\Ktot}{2}\cdot(\mathbf r_i+\mathbf r_j)}
    e^{-i\qq\cdot(\mathbf r_j-\mathbf r_i)}
    \ket{i,j}.
    \label{eq:two_magnon_Kq}
\end{equation}
Since the two flipped spins are indistinguishable, 
\begin{equation}
  \ket{\Ktot,\qq}_{\mathrm R}=\ket{\Ktot,-\qq}_{\mathrm R},
  \label{eq:evenKq}
\end{equation}
the wave function describing the two flipped \(S=1/2\) is even under \(\qq\to-\qq\). 

A general two-magnon wave function at fixed total momentum \(\Ktot\) is then written as
\begin{equation}
    \ket{\Psi_{\Ktot}}
    =
    \sum_{\qq}
    \psi_{\Ktot}^{\mathrm R}(\qq)
    \ket{\Ktot,\qq}_{\mathrm R}.
    \label{eq:two_magnon_wavefunction}
\end{equation}
Throughout this section, \(\varepsilon_{\qq}\) and \(E_2\) denote
zero-field excitation energies measured relative to the energy of the
fully polarized state--a magnetic field \(h\) shifts an \(n\)-magnon
excitation energy by \(nh\).
The Schr\"odinger equation for the momentum-space wave function is
\begin{equation}
    \left(
    E_2
    -
    \varepsilon_{\frac{\Ktot}{2}-\qq}
    -
    \varepsilon_{\frac{\Ktot}{2}+\qq}
    \right)
    \psi_{\Ktot}^{\mathrm R}(\qq)
    =
    \frac{1}{N}
    \sum_{\pp}
    V_{\Ktot}(\qq,\pp)
    \psi_{\Ktot}^{\mathrm R}(\pp),
    \label{eq:two_magnon_schrodinger}
\end{equation}
with interaction kernel
\begin{align}
    V_{\Ktot}(\qq,\pp)
    &=
    \frac{1}{2}
    \left(
    \mathcal{J}_{\qq-\pp}
    +\mathcal{J}_{\qq+\pp}
    -\mathcal{J}_{\frac{\Ktot}{2}-\pp}
    -\mathcal{J}_{\frac{\Ktot}{2}+\pp}
    \right)
    \nonumber\\
    &=
    \varepsilon_{\qq-\pp}
    +\varepsilon_{\qq+\pp}
    -\varepsilon_{\frac{\Ktot}{2}-\pp}
    -\varepsilon_{\frac{\Ktot}{2}+\pp},
    \label{eq:kernel}
\end{align}
where the second equality follows from
\(\varepsilon_{\qq}=S(\mathcal J_{\qq}-\mathcal J_0)\) with \(S=1/2\) (the momentum-independent \(\mathcal J_0\) terms cancel).
We note that this kernel is the transpose of the more common convention
in which the hard-core terms depend on \(\qq\) rather than on \(\pp\)
\cite{Zhitomirsky_2010,jiang2023where}.
The two conventions give the same bound-state energies.

Using the lattice harmonics defined in Appendix~\ref{sec:trilatt} and the decomposition Eq.~\eqref{eq:decomposeChi}, the kernel is separable as
\begin{equation}
    V^{(\mathrm{H})}_{\Ktot}(\qq,\pp) = \sum_{\mu}V_{\mu} \bar\gamma_{\pp}^{\mu} R_{\mu}\left(\qq,\tfrac{\Ktot}{2}\right),
    \label{eq:separable_kernel}
\end{equation}
with
\begin{equation}
    R_{\mu}\left(\qq,\tfrac{\Ktot}{2}\right)
    =\gamma^{\mu}_{\qq}-\gamma^{\mu}_{\tfrac{\Ktot}{2}},
    \label{eq:RmuqK}
\end{equation}
and
\begin{equation}
V_{\mu}
=
\begin{cases}
6J_n, & \mu = (n)\;, \\
\frac{3}{2}J_n, & \mu = (n+) \;, \\
\frac{3}{2}J_n, & \mu = (n-) \;, \\
\end{cases}
\qquad n=1,2,3 .
\label{eq:Vmu}
\end{equation}
Here $\mu$ enumerates the lattice harmonics defined in Eq.~\eqref{eq:C6harmonics} and \(\bar\gamma_{\pp}^{\mu}\) are complex conjugates of \(\gamma_{\pp}^{\mu}\) .
This form is the triangular-lattice analog of the partial-wave decomposition in the square-lattice problem~\cite{Zhitomirsky_2010,jiang2023where,gallegos_chernyshev}.

\subsection{Effect of the scalar-chirality term - revised}
\label{sec:chirality}

We now include the scalar-chirality interaction in the two-magnon problem. As shown in Sec.~\ref{sec:chi_one_magnon}, \(\hat H_{\mathrm{Chir}}\) has no matrix elements within the one-magnon subspace and therefore does not modify the one-magnon dispersion \(\varepsilon_{\qq}\). Its first nontrivial effect appears in the two-magnon sector, where it contributes directly to the interaction kernel \(V_{\Ktot}(\qq,\pp)\).

The general chirality contribution
\(V_{\Ktot}^{(\chi)}(\qq,\pp)\) is derived in
Appendix~\ref{app:chirality_kernel}, the final form of the contribution expressed with the complex \(C_6\) harmonics reads
\begin{align}
V_{\Ktot}^{(\chi)}(\qq,\pp)
&=
6\sqrt{3}J_{\chi}
\bigg[
\gamma_{\frac{\Ktot}{2}}^{(1)}
\left(
\gamma_{\pp}^{(1+)*}
\gamma_{\qq}^{(1+)}
-
\gamma_{\pp}^{(1-)*}
\gamma_{\qq}^{(1-)}
\right)
\nonumber\\
&\hspace{1.7cm}
+
\gamma_{\frac{\Ktot}{2}}^{(1+)}
\left(
\gamma_{\pp}^{(1)*}
\gamma_{\qq}^{(1-)}
-
\gamma_{\pp}^{(1+)*}
\gamma_{\qq}^{(1)}
\right)
\nonumber\\
&\hspace{1.7cm}
+
\gamma_{\frac{\Ktot}{2}}^{(1-)}
\left(
\gamma_{\pp}^{(1-)*}
\gamma_{\qq}^{(1)}
-
\gamma_{\pp}^{(1)*}
\gamma_{\qq}^{(1+)}
\right)
\bigg].
\label{eq:app_int_chi_complex_harmonics}
\end{align}

At finite total momentum it mixes
the symmetric and chiral first-neighbor lattice harmonics. At
\(\Ktot=0\), however, the mixing terms vanish, and the
scalar-chirality interaction reduces to opposite shifts of the two
chiral components of the \(\Etwo\) doublet, as made explicit by the
first-neighbor couplings \(V_{1\pm}\) in Eq.~\eqref{eq:V1pm}.

Time reversal maps \(J_\chi\to -J_\chi\) and interchanges the two
chiral sectors. Consequently, at \(\Ktot=0\),
\begin{equation}
    E_{2,+}(J_\chi)
    =
    E_{2,-}(-J_\chi).
\end{equation}

The full two-magnon interaction kernel is
\begin{equation}
    V_{\Ktot}(\qq,\pp)
    =
    V_{\Ktot}^{(\mathrm H)}(\qq,\pp)
    +
    V_{\Ktot}^{(\chi)}(\qq,\pp),
    \label{eq:full_two_magnon_kernel}
\end{equation}
where \(V_{\Ktot}^{(\mathrm H)}\) is the Heisenberg contribution given
in Eq.~\eqref{eq:kernel}.

\subsection{Gap equation for the bound-state criterion}
\label{sec:channel_decomposition}

\subsubsection{Binding energy}

A bound state at fixed total momentum \(\Ktot\) occurs when its energy
\(E_2(\Ktot)\) lies below the lower edge of the two-magnon continuum,
\begin{equation}
    \Econtmin(\Ktot)
    =
    \min_{\qq}
    \left[
        \varepsilon_{\frac{\Ktot}{2}+\qq}
        +
        \varepsilon_{\frac{\Ktot}{2}-\qq}
    \right].
\end{equation}
We define the corresponding two-magnon binding energy by
\begin{equation}
    E_2(\Ktot)
    =
    \Econtmin(\Ktot)
    -
    \Delta_2(\Ktot),
    \qquad
    \Delta_2(\Ktot)>0.
    \label{eq:binding_energy}
\end{equation}
The instability of the fully polarized state is determined by the
lowest energy per flipped spin.
As discussed in connection with
Eq.~\eqref{eq:critical_field_one_magnon}, the one-magnon critical field
is
\begin{equation}
    h_c^{(1)}
    =
    -\min_{\qq}\varepsilon_{\qq},
\end{equation}
while the two-magnon critical field is
\begin{equation}
    h_c^{(2)}
    =
    -\frac{1}{2}
    \min_{\Ktot}E_2(\Ktot).
    \label{eq:critical_field_two_magnon}
\end{equation}
A two-magnon instability preempts the one-magnon instability when
\begin{equation}
    h_c^{(2)}>h_c^{(1)},
\end{equation}
or, equivalently,
\begin{equation}
    \min_{\Ktot}E_2(\Ktot)
    <
    2\min_{\qq}\varepsilon_{\qq}.
    \label{eq:two_magnon_preemption}
\end{equation}

\subsubsection{Gap equation}

We now restrict the gap-equation analysis to total momentum
\(\Ktot=0\), corresponding to the \(\Gamma\) point.
Since the one-magnon dispersion is inversion symmetric,
\(\varepsilon_{\qq}=\varepsilon_{-\qq}\), the lower edge of the
two-magnon continuum at \(\Ktot=0\) is
\begin{equation}
    \Econtmin(0)
    =
    2\varepsilon_{\min},
    \qquad
    \varepsilon_{\min}
    =
    \min_{\qq}\varepsilon_{\qq}.
\end{equation}
We suppress the \(\Ktot=0\) argument from \(E_2\) and \(\Delta_2\)
from this point on.
The critical field associated with a \(\Ktot=0\) bound state is then
\begin{equation}
    h_c^{(2)}(0)
    =
    h_c^{(1)}
    +
    \frac{\Delta_2}{2}.
    \label{eq:hc2_binding}
\end{equation}
Thus, any finite binding energy at \(\Ktot=0\) causes this two-magnon
state to become gapless at a higher field than the one-magnon mode.
Provided that no bound state of three or more magnons has a still
higher critical field, this signals a spin-nematic instability.
The restriction to \(\Ktot=0\) does not exclude bound states at finite
total momentum; these are examined separately using momentum-resolved
exact diagonalization below.

For \(\Ktot=0\), the scalar-chirality contribution reduces to opposite
shifts of the first-neighbor chiral couplings,
\begin{subequations}
\begin{align}
    V_{1+}
    &=
    \frac{3}{2}J_1
    +
    6\sqrt{3}J_\chi,
    \\
    V_{1-}
    &=
    \frac{3}{2}J_1
    -
    6\sqrt{3}J_\chi,
\end{align}
\end{subequations}
while the fully symmetric \(\Aone\) channel remains unchanged.
The interaction kernel can therefore be written as
\begin{equation}
    V_{0}(\qq,\pp)
    =
    \sum_{\mu}
    V_{\mu}\,
    \bar\gamma_{\pp}^{\mu}
    R_{\mu}(\qq,0),
\end{equation}
where
\begin{equation}
    R_{\mu}(\qq,0)
    =
    \gamma_{\qq}^{\mu}
    -
    \gamma_{\mathbf 0}^{\mu}.
\end{equation}
Substituting this separable form into
Eq.~\eqref{eq:two_magnon_schrodinger} gives
\begin{equation}
    \left(
        E_2-2\varepsilon_{\qq}
    \right)
    \psi^{\mathrm R}_{0}(\qq)
    =
    \frac{1}{N}
    \sum_{\pp,\mu}
    V_{\mu}
    R_{\mu}(\qq,0)
    \bar\gamma_{\pp}^{\mu}
    \psi^{\mathrm R}_{0}(\pp).
    \label{eq:psi_separable}
\end{equation}
We introduce the partial-wave amplitudes
\begin{equation}
    S_{\mu}
    =
    \frac{1}{N}
    \sum_{\pp}
    \bar\gamma_{\pp}^{\mu}
    \psi^{\mathrm R}_{0}(\pp).
    \label{eq:Smu_def}
\end{equation}
Equation~\eqref{eq:psi_separable} then yields
\begin{equation}
    \psi^{\mathrm R}_{0}(\qq)
    =
    \sum_{\mu}
    \frac{
        V_{\mu}
        R_{\mu}(\qq,0)
    }{
        E_2-2\varepsilon_{\qq}
    }
    S_{\mu}.
    \label{eq:selfconsistency_wavefunction}
\end{equation}
Substituting this expression back into Eq.~\eqref{eq:Smu_def} gives
a closed homogeneous system for the partial-wave amplitudes,
\begin{equation}
    S_{\nu}
    =
    \sum_{\mu}
    \mathcal{I}_{\nu\mu}(E_2)
    S_{\mu},
    \label{eq:S_mu_equation}
\end{equation}
with
\begin{equation}
    \mathcal{I}_{\nu\mu}(E_2)
    =
    \frac{1}{N}
    \sum_{\pp}
    \frac{
        V_{\mu}
        \bar\gamma_{\pp}^{\nu}
        R_{\mu}(\pp,0)
    }{
        E_2-2\varepsilon_{\pp}
    }.
    \label{eq:I_numu_general}
\end{equation}
A nontrivial solution exists when
\begin{equation}
    \det\!\left[
    \mathcal{I}(E_2)-\mathbf{1}
    \right]
    =
    0\,.
    \label{eq:secular_condition}
\end{equation}
This is the gap equation for the two-magnon bound state.
At \(\Ktot=0\), the gap equation separates into three independent \(3\times3\) sectors. At \( J_\chi = 0\), these comprise the fully symmetric \(\Aone\) sector and the two degenerate components of the \(D_6\) \(\Etwo\) doublet. A finite \( J_\chi \) reduces the point-group symmetry to \( C_6 \), under which the latter separate into the two chiral sectors \(\Etwop\) and \(\Etwom\).

For the \(\Aone\) sector, the matrix is
\begin{subequations}
\label{eq:gapA1}
\begin{equation}
\begin{pmatrix}
    \mathcal{I}_{1,1}
    &
    \mathcal{I}_{1,2}
    &
    \mathcal{I}_{1,3}
    \\
    \mathcal{I}_{2,1}
    &
    \mathcal{I}_{2,2}
    &
    \mathcal{I}_{2,3}
    \\
    \mathcal{I}_{3,1}
    &
    \mathcal{I}_{3,2}
    &
    \mathcal{I}_{3,3}
\end{pmatrix},
\label{eq:three_channel_1}
\end{equation}
with matrix elements
\begin{equation}
    \mathcal{I}_{n,m}(E_2)
    =
    \frac{1}{N}
    \sum_{\pp}
    \frac{
        V_m
        \gamma_{\pp}^{(n)}
        \left[
            \gamma_{\pp}^{(m)}-1
        \right]
    }{
        E_2-2\varepsilon_{\pp}
    },
    \label{eq:I_nm}
\end{equation}
and couplings
\begin{equation}
    (V_1,V_2,V_3)
    =
    (6J_1,6J_2,6J_3).
\end{equation}
\end{subequations}
For the \(\Etwopm\) chiral sectors, the matrix elements are
\begin{subequations}
\label{eq:gapE2}
\begin{equation}
    \mathcal{I}_{n\pm,m\pm}(E_2)
    =
    \frac{1}{N}
    \sum_{\pp}
    \frac{
        V_{m\pm}
        \bar\gamma_{\pp}^{(n\pm)}
        \gamma_{\pp}^{(m\pm)}
    }{
        E_2-2\varepsilon_{\pp}
    },
    \label{eq:I_pm}
\end{equation}
with
\begin{align}
    V_{1\pm}
    &=
    \frac{3}{2}J_1
    \pm
    6\sqrt{3}J_\chi,
    \label{eq:V1pm}
    \\
    V_{2\pm}
    &=
    \frac{3}{2}J_2,
    \\
    V_{3\pm}
    &=
    \frac{3}{2}J_3.
\end{align}
\label{eq:Vs}
\end{subequations}
Thus, the scalar-chirality interaction lifts the degeneracy between
the \(\Etwop\) and \(\Etwom\) sectors by shifting their
first-neighbor couplings in opposite directions.
For \(J_\chi\neq0\), the one-magnon dispersion, and hence the
denominator in Eq.~\eqref{eq:I_numu_general}, remains unchanged.
The scalar chirality enters only through the first-neighbor couplings
\(V_{1\pm}\), Eq.~\eqref{eq:V1pm}, and therefore only through the
corresponding column of each \(\Etwopm\) block of
\(\mathcal{I}-\mathbf{1}\).
Since the determinant is linear in each column separately, for fixed
\(E_2\) and \((J_1,J_2,J_3)\) the secular equation is linear in
\(J_\chi\).
It can therefore be solved directly for the value of \(J_\chi\)
corresponding to any prescribed two-magnon energy \(E_2\).
At zero field, for example, the boundary at which the
\(\Ktot=0\) two-magnon excitation becomes gapless is obtained by
setting \(E_2=0\) in the secular equation.

The same construction can be used to probe the onset of two-magnon
binding.
For \(\Ktot=0\), we write
\begin{equation}
    E_2
    =
    2\varepsilon_{\min}
    -
    \Delta_2,
    \qquad
    \varepsilon_{\min}
    =
    \min_{\qq}\varepsilon_{\qq},
    \label{eq:E2_binding_gap}
\end{equation}
with \(\Delta_2>0\).
For a prescribed \(\Delta_2\), the secular equation can again be
solved directly for \(J_\chi\).
The exact binding threshold corresponds formally to the limit
\(\Delta_2\to0^+\).
Directly setting \(\Delta_2=0\), however, makes the denominator in
Eq.~\eqref{eq:I_numu_general} singular at the one-magnon minima and is
therefore inconvenient for numerical evaluation.
Instead, we evaluate the secular condition at small finite values of
\(\Delta_2\) and plot the resulting contours in parameter space.
These contours locate where a bound state with the prescribed binding
energy exists and approach the binding threshold as
\(\Delta_2\to0^+\).

\begin{figure*}[t]
    \centering
    \includegraphics[width=0.7\textwidth]{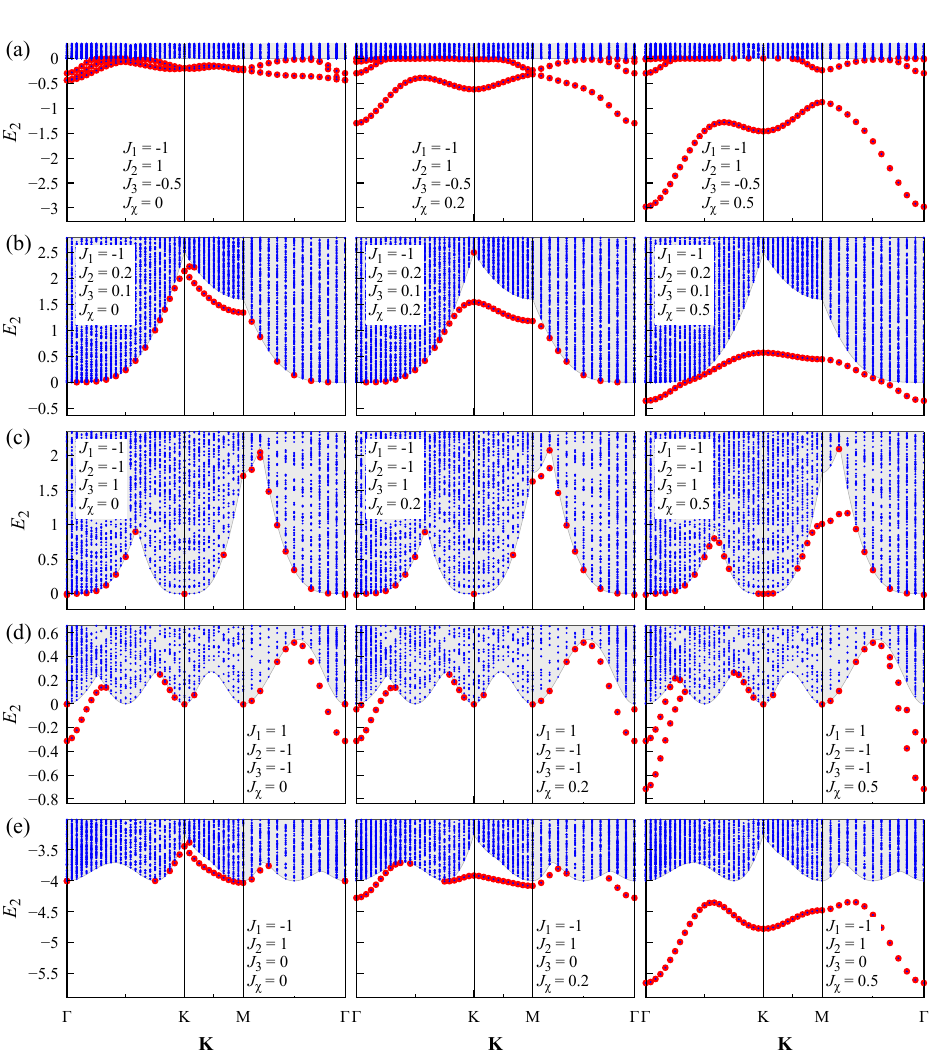}
\caption{
Exact-diagonalization spectra in the two-magnon sector for an
\(N=3\times24^2=1728\) cluster at selected \((J_1,J_2,J_3)\).
The three columns correspond to \(J_\chi=0\), \(0.2\), and \(0.5\),
while rows (a)--(e) correspond to the representative exchange parameters discussed in the text.
The total momentum \(\Ktot\) follows the
\(\Gamma\)-\(K\)-\(M\)-\(\Gamma\) path.
Gray shading denotes the two-magnon continuum, blue points the
finite-size two-magnon energies, and red circles levels below the
lower continuum edge.
Since the one-magnon dispersion is independent of \(J_\chi\), the
continuum is unchanged. Finite \(J_\chi\) selectively lowers
part of the two-magnon spectrum; at \(\Gamma\), the gap-equation
analysis identifies this effect with the splitting of the two opposite
chiral sectors.
(a) Fine-tuned point realizing the
\(\mathcal{M}_{\Btwo}\) nodal-line minimum manifold; at \(J_\chi=0\),
the flat lower continuum edge reflects the dispersionless one-magnon
minima along high-symmetry lines.
(b) Point where the Lifshitz boundary meets the
\(\mathcal{M}_{\Aone}\) manifold.
(c) Point where the ferromagnetic phase boundary meets the \(K\) and
\(\Gamma K\) regions. Finite-size levels lie slightly below the
continuum already at \(J_\chi=0\), while a clearly separated
\(\Gamma\)-point bound-state branch develops only for
\(J_\chi\gtrsim0.66\), as shown below.
(d) Point exhibiting one of the largest binding energies in the scanned antiferromagnetic-\(J_1\) regime. At \(J_\chi=0\), a well-developed
\(\Gamma\)-point bound state is present; the gap equation identifies
it as an \(\Aone\) state. Finite \(J_\chi\) lowers one of the chiral
\(\Etwo\) components.
(e) Example where the lowest level below the continuum occurs at \(M\) point for \(J_\chi=0\), and changes to \(\Gamma\)-point when increasing \(J_\chi\).
}
\label{fig:ED1728spectra}
\end{figure*}

\begin{figure}[tb]
    \centering
    \includegraphics[width=0.9\linewidth]{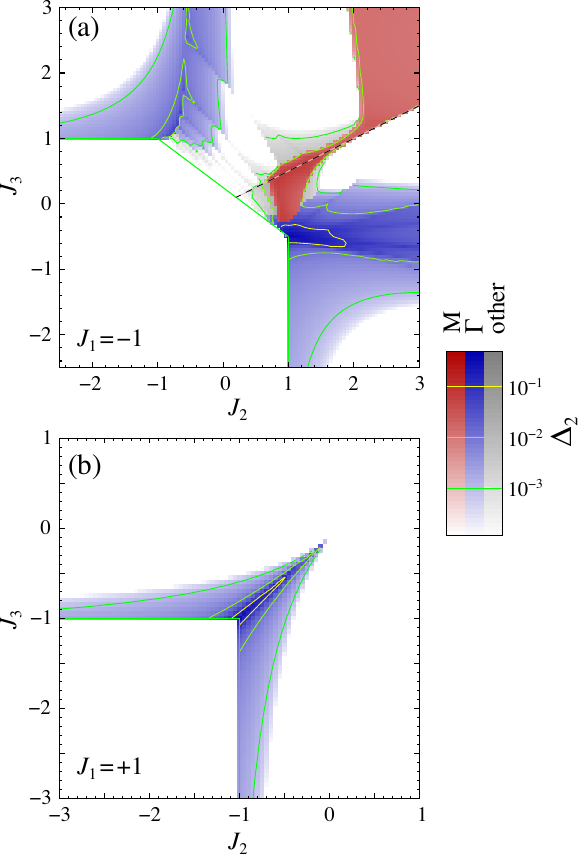}
    \caption{%
    Scan of the total momentum \(\Ktot\) and binding energy \(\Delta_2\)
    of the lowest-energy two-magnon bound state in the \(J_2\)-\(J_3\)
    parameter space for \(J_\chi = 0\), with (a) \(J_1=-1\) and (b) \(J_1=+1\).
    The binding energy is obtained from Eq.~\eqref{eq:binding_energy}
    using exact-diagonalization energies for an
    \(N=3\times24^2=1728\)-site cluster.
    The lowest-energy bound state occurs predominantly at the
    \(\Gamma\) point, but regions with \(\Ktot=M\) and with
    incommensurate total momentum also appear.
    The latter are found mainly where the one-magnon dispersion has
    incommensurate minima along the \(\Gamma K\) or \(\Gamma M\)
    directions.
    The largest binding energies in the scanned regions occur at
    \((J_1,J_2,J_3)=(-1,1,-1/2)\) in panel (a) and
    \((J_1,J_2,J_3)=(1,-1,-1)\) in panel (b).
    The thin green lines mark the boundary of the ferromagnetic phase,
    while the dashed black-and-white line indicates the
    \(\mathcal{M}_{\Aone}\) manifold.
    }
    \label{fig:ED1728}
\end{figure}

\begin{figure}[tb]
    \centering
    \includegraphics[width=0.95\linewidth]{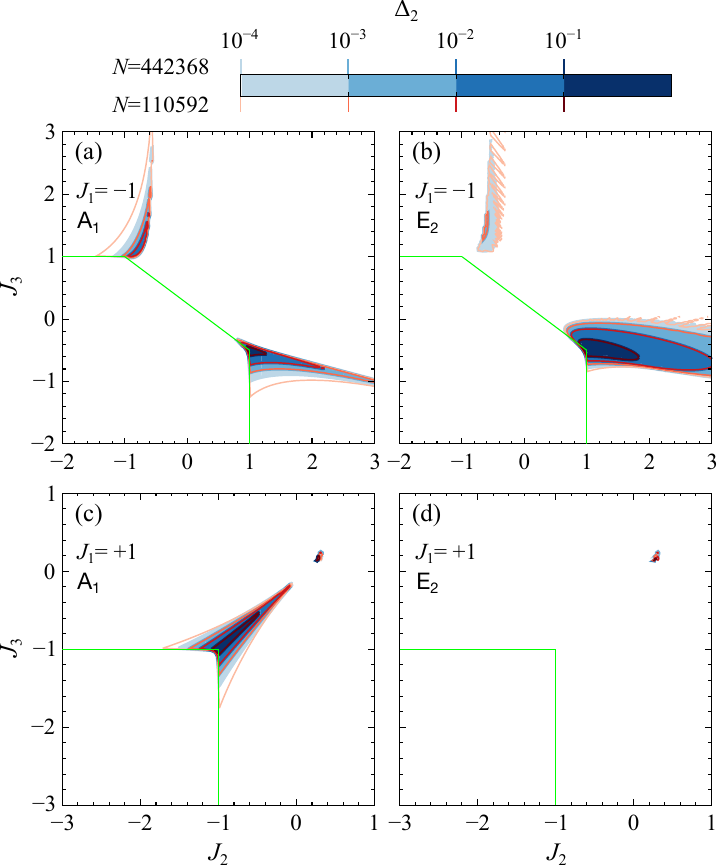}
\caption{%
Contours of the two-magnon binding energy $\Delta_2$
obtained at $J_\chi=0$ from the $\Ktot=0$ gap equation,
Eq.~\eqref{eq:secular_condition}, using momentum meshes
containing $3\times192^2=110\,592$ (orange contours) and
$3\times384^2=442\,368$ (blue contours) points.
The blue shading indicates the binding energy, with
darker shades corresponding to stronger binding.
Panels (a,c) correspond to the $\Aone$ channel and
(b,d) to the $\Etwo$ channel, with $J_1=-1$ in (a,b)
and $J_1=+1$ in (c,d).
Unlike Fig.~\ref{fig:ED1728}, which presents the
lowest bound states at arbitrary total momentum,
these calculations resolve the $\Ktot=0$ states
into symmetry channels using substantially finer
momentum meshes.
The contours are well converged for
$\Delta_2\gtrsim10^{-3}$, while finite-size corrections
remain visible at $\Delta_2=10^{-4}$, particularly
in the incommensurate regions.
These corrections account for apparent differences
near the weak-binding boundaries.
The largest binding in the $\Aone$ channel occurs at $(J_1,J_2,J_3)=(1,-1,-1)$, while that in the $\Etwo$ channel occurs at $(-1,1,-1/2)$.
The green lines mark the boundary of the ferromagnetic phase.
}
\label{fig:gaps_partial}
\end{figure}

\begin{figure*}[tb]
    \centering
    \includegraphics[width=0.85\linewidth]{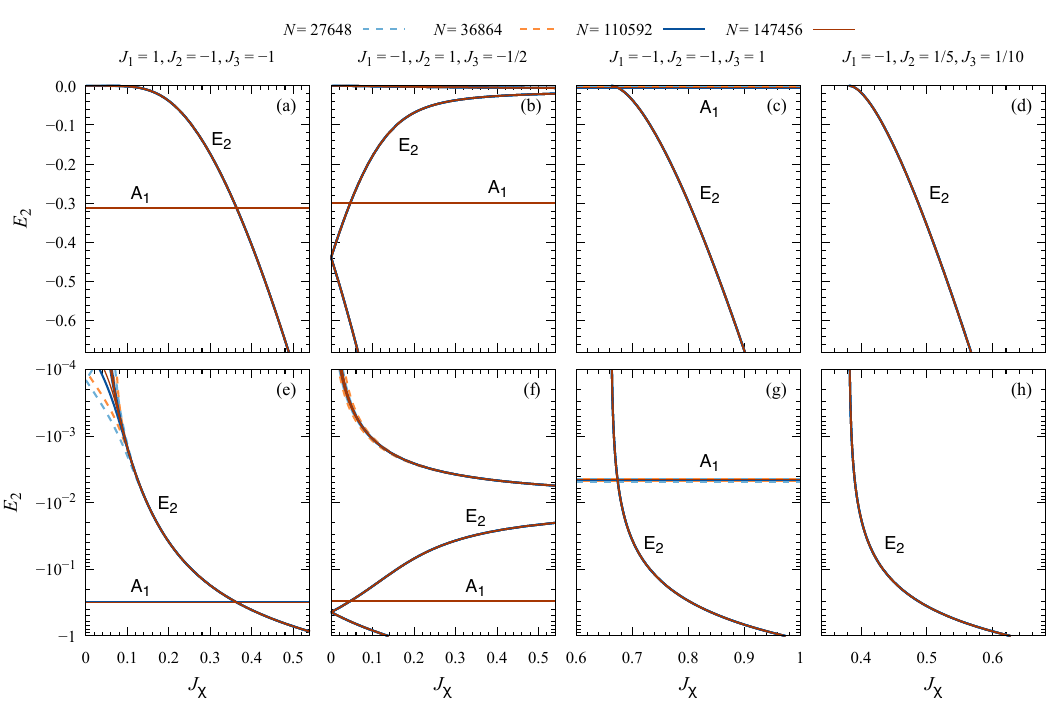}
    \caption{%
    Evolution of the two-magnon bound-state energies as a function of scalar-chirality \(J_\chi\) at the \(\Gamma\)-point, measured from the fully polarized energy $E_0$ for four sets of exchange parameters (columns).
Panels (a)--(d) show the energies on a linear scale; panels (e)--(h)
show the same data on a logarithmic scale in \(|E_2|\), which
resolves the weakly bound states near threshold.
Note that the \(J_\chi\) ranges differ between columns. Only
\(J_\chi\ge 0\) is shown; the spectra are symmetric under
\(J_\chi\to-J_\chi\) with \(\Etwop\leftrightarrow\Etwom\).
The energies are obtained from the gap equation in the \(\Aone\) and
\(\Etwo\) sectors, Eqs.~(\ref{eq:gapA1}) and (\ref{eq:gapE2}).
The \(\Aone\) branches (horizontal lines) are independent of
\(J_\chi\), whereas scalar chirality splits the initially twofold-degenerate
\(\Etwo\) doublet into the opposite chiral components \(\Etwop\) and
\(\Etwom\).
The Brillouin-zone integrals were evaluated on momentum meshes with
\(192^2=36864\) and \(384^2=147456\) points, and
\(3\times96^2=27648\) and \(3\times192^2=110592\) points. Results on symmetry-preserving shifted meshes of the same
sizes, are drawn in the same style.
The two finest meshes agree within the linewidth for
\(|E_2|\gtrsim10^{-3}\); the coarser and/or shifted meshes deviate visibly only
close to threshold, i.e., at small \(J_\chi\) in (e) and (f), and for
the shallow \(\Aone\) state in (g).
In (a,e) and (c,g), an \(\Aone\) bound state is present at
\(J_\chi=0\), and a chiral \(\Etwo\) state becomes the lowest
\(\Gamma\)-point bound state above \(J_\chi\approx0.36\) and
\(J_\chi\approx0.67\), respectively.
In (b,f), the \(\Etwo\) doublet is bound below the \(\Aone\) state
already at \(J_\chi=0\), and its splitting into chiral components is
directly visible.
In (d,h), no \(\Gamma\)-point bound state is present at
\(J_\chi=0\), and finite scalar chirality (\(J_\chi\gtrsim0.38\)) is
required to produce binding at \(\Gamma\).
    }
    \label{fig:gap_splitting}
\end{figure*}

\begin{figure}[tb]
    \centering
    \includegraphics[width=0.85\linewidth]{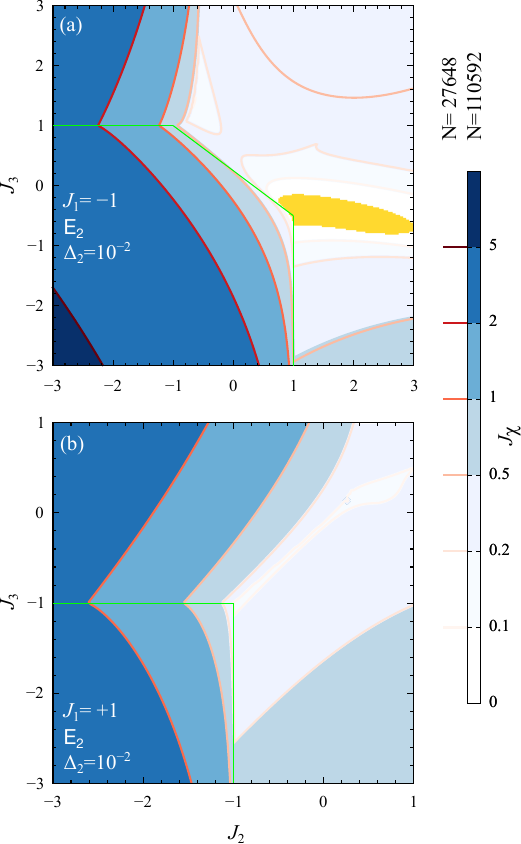}
    \caption{%
    Contours of constant two-magnon binding energy
    \(\Delta_2=10^{-2}\) for several values of \(|J_\chi|\).
    For both signs of \(J_1\), scalar chirality substantially enlarges
    the parameter region supporting a bound state in the chiral
    \(\Etwo\) channel.
    The colored regions are obtained using a momentum mesh of
    \(N=3\times192^2=110592\) points in the Brillouin zone, while the
    red contours show the corresponding boundaries for
    \(N=3\times96^2=27648\).
    Their near coincidence indicates that finite-size corrections are
    small at this binding-energy scale.
    The orange region denotes the bound-state regime already present
    at \(J_\chi=0\).
    The green lines mark the boundary of the ferromagnetic phase.
    }
    \label{fig:Jchi_contours}
\end{figure}

\section{Results}
\label{sec:results}

We now apply the two-magnon formulation developed above to the triangular-lattice \(J_1\)-\(J_2\)-\(J_3\) model. We use two complementary approaches. Momentum-resolved exact diagonalization in the two-magnon sector gives access to the lowest states at all total momenta \(\Ktot\) allowed by the finite cluster, while the \(\Ktot=0\) gap equation resolves the corresponding bound states into the \(\Aone\) and chiral \(\Etwopm\) partial-wave channels. Together, these calculations allow us to distinguish the momentum-space structure of the two-magnon spectrum from the symmetry mechanism by which scalar chirality enhances binding. 
We use periodic clusters preserving the full $D_6$ point-group symmetry and momentum meshes preserving at least threefold rotational symmetry as described in Appendix~\ref{sec:trilatt}.

To achieve sufficiently high momentum-space resolution, we considered large (up to 1728 sites) finite clusters and constructed the two-magnon Hilbert space directly in a translationally invariant basis, thereby block-diagonalizing the Hamiltonian into sectors of fixed total momentum. We then obtained the lowest eigenvalue in each momentum sector using the built-in eigensolver in Wolfram Mathematica, with the Arnoldi method specified. We additionally implemented an independent Lanczos algorithm, but found no significant computational advantage over the built-in solver.

Figure~\ref{fig:ED1728spectra} shows representative two-magnon spectra on an \(N=3\times24^2=1728\)-site cluster. The gray region denotes the two-magnon continuum constructed from the one-magnon dispersion, while the discrete levels below its lower edge are candidate bound states. For levels that lie only slightly below the finite-size continuum, this identification must be treated with caution, since the separation may vanish in the thermodynamic limit. By contrast, the strongly separated levels seen in several representative parameter sets provide clear evidence for robust two-magnon binding.

\subsection{Results for \(J_\chi=0\): no chiral terms}

At \(J_\chi=0\), the spectra already reveal a strong dependence of the binding on the structure of the one-magnon minima. The most pronounced example for \(J_1=-1\) occurs at
\begin{equation}
    (J_1,J_2,J_3)=(-1,1,-1/2),
\end{equation}
where the \(\mathcal{M}_{\Btwo}\) nodal-line minimum manifold is realized. Here the lower edge of the two-magnon continuum is anomalously flat, and a strongly bound \(\Etwo\)-type state occurs at \(\Gamma\). As shown below by the symmetry-resolved gap equation, this \(\Gamma\)-point state belongs to the \(\Etwo\) sector. In contrast, at
\begin{equation}
    (J_1,J_2,J_3)=(-1,1/5,1/10),
\end{equation}
where the \(\mathcal{M}_{\Aone}\) manifold meets the ferromagnetic Lifshitz boundary, no comparably separated bound state is resolved. Thus, a soft one-magnon dispersion alone does not determine the strength of the two-magnon instability; the detailed geometry and degeneracy of its minimum manifold are important.

The exact spectra also show that the minimum of the two-magnon bound-state dispersion need not occur at \(\Gamma\). A particularly clear example is
\begin{equation}
    (J_1,J_2,J_3)=(-1,1,0),
\end{equation}
where the lowest energy level below the two-magnon continuum occurs at \(M\), while the \(\Gamma\)-point binding is much weaker. The momentum-resolved scan in Fig.~\ref{fig:ED1728} summarizes all these findings throughout the \(J_2\)-\(J_3\) plane for both signs of \(J_1\). The lowest two-magnon states occur predominantly at \(\Ktot=\Gamma\), but finite regions with \(\Ktot=M\), as well as incommensurate total momenta, are also present. The incommensurate regions occur mainly where the one-magnon dispersion itself has incommensurate minima along the \(\Gamma K\) or \(\Gamma M\) directions. A complete analytic treatment of these regions would therefore require extending the partial-wave gap equation to finite \(\Ktot\).

Within the parameter regions studied here, the largest binding energies occur near two special points. For \(J_1=-1\), the strongest binding is found at the \(\mathcal{M}_{\Btwo}\) point \((-1,1,-1/2)\). For \(J_1=+1\), the largest binding in the scanned window occurs near \((1,-1,-1)\). These are multicritical points where several phases meet. Representative values are collected in Table~\ref{tab:gap-values}.
Conversely, the gap \(\Delta_2\) at \((J_1,J_2,J_3)=(-1,-1,1)\) is small, while no resolved gap is found at \((-1,1/5,1/10)\). These comparisons show that frustration or one-magnon softening alone is not sufficient to produce strong two-magnon binding; the geometry and degeneracy of the low-energy one-magnon manifold play a decisive role.

\begin{table}[t]
\caption{%
Energies for selected \(J_1,J_2,J_3\) parameter sets at \(J_\chi=0\), obtained from exact diagonalization on the \(N=3\times24^2=1728\)-site cluster and from the \(\Ktot=0\) gap equations in the thermodynamic limit. The \(\Etwo\) levels are twofold degenerate at \(J_\chi=0\). The \(\Econtmin(\Ktot)\) denotes the lower edge of the two-magnon continuum.
}
\label{tab:gap-values}
\begin{ruledtabular}
\begin{tabular}{llccccc}
\multirow{2}{*}{\((J_1,J_2,J_3)\)}
&
\multirow{2}{*}{Ch.}
&
\multirow{2}{*}{\(\Econtmin(\Ktot)\)}
&
\multicolumn{2}{c}{\(\Delta_2\)} \\
&
&
&
\multicolumn{1}{c}{\(N=1728\)}
&
\multicolumn{1}{c}{\(N\to \infty\)} 
\\
\hline
\multirow{2}{*}{\((1,-1,-1)\)}
& \(\Aone\)
& \multirow{2}{*}{\(0\)}
& \(0.312424\)
& \(0.312237\)
\\
& \(\Etwo\)
&
& \(0.001826\)
& \(0\)
\\
\\
\multirow{2}{*}{\((-1,1,-\frac{1}{2})\)}
& \(\Aone\)
&
& \(0.299375\)
& \(0.299375\)
\\
& \(\Etwo\) 
& \multirow{2}{*}{\(0\)}
& \(0.438620\)
& \(0.438620\)
\\
\\
\((-1,-1,1)\)
& \(\Aone\)
& \(0\)
& \(0.017163\)
& \(0.004493\)
\\
\\
\((-1,\frac{1}{5},\frac{1}{10})\)
& \(\Aone\)
& \(0\)
& \(0\)
& \(0\)
\\
\\
\multirow{3}{*}{\((-1,1,0)\)}
& \(M\)
& \multirow{3}{*}{\(-4\)}
& \(0.032352\)
& \(\cdots\)
\\
& \(\Gamma\; \Aone\)
& 
& \(0\)
& \(0\)
\\
& \(\Gamma\; \Etwo\)
& 
& \(0.002146\)
& \((2.3\pm 0.1)\times 10^{-5}\)
\\
\end{tabular}
\end{ruledtabular}
\end{table}

To identify the symmetry of the \(\Gamma\)-point bound states, we next solve the \(\Ktot=0\) gap equation. Figure~\ref{fig:gaps_partial} shows contours of fixed binding energy in the \(\Aone\) and \(\Etwo\) sectors at \(J_\chi=0\). The calculations use momentum meshes up to \(3\times384^2\) points. Contours corresponding to binding energies \(\Delta_2\gtrsim10^{-3}\) are already well converged, whereas the \(\Delta_2=10^{-4}\) contours still show visible finite-size corrections. These corrections are particularly strong in the incommensurate regions, where the discrete momentum mesh samples the one-magnon minima irregularly. We therefore use the larger-gap contours to identify the robust structure of the bound-state regions and do not attach significance to fine details of the smallest-gap boundaries.

The partial-wave calculation provides symmetry information not contained in the momentum-resolved spectra alone. For \(J_1=-1\), the strongest \(\Gamma\)-point binding occurs in the \(\Etwo\) sector around the \(\mathcal{M}_{\Btwo}\) point. For \(J_1=+1\), the largest robust region in the scanned parameter range instead occurs in the fully symmetric \(\Aone\) channel near \((1,-1,-1)\). The enhancement near \(\mathcal{M}_{\Btwo}\) is consistent with the unusually large low-energy phase space generated by the nodal-line minimum of the one-magnon dispersion. Thus, the symmetry of the leading \(\Gamma\)-point bound state is not fixed by lattice symmetry alone, but depends sensitively on which frustrated manifold controls the low-energy one-magnon dispersion.

\subsection{Results for finite \(J_\chi\)}

We now turn to the main effect of scalar chirality. Figure~\ref{fig:gap_splitting} follows the \(\Gamma\)-point two-magnon energies as a function of \(J_\chi\) for representative exchange parameters. The distinction between the two symmetry sectors is immediate. The \(\Aone\) bound-state energies are independent of \(J_\chi\), whereas the two components of the \(\Etwo\) doublet, degenerate at \(J_\chi=0\), split into the opposite chiral sectors \(\Etwop\) and \(\Etwom\). One component is driven to lower energy while the other is pushed upward. Reversing \(J_\chi\) interchanges the two components, in accordance with
\begin{equation}
    E_{2,+}(J_\chi)=E_{2,-}(-J_\chi).
\end{equation}
Because the one-magnon continuum is unchanged, the energy of the lower of the two chiral \(\Etwo\) branches, and hence its binding relative to the continuum, is even under \(J_\chi\to-J_\chi\), while the chirality of the selected branch is reversed.

The examples in Fig.~\ref{fig:gap_splitting} illustrate two qualitatively distinct consequences of this splitting. When a bound state is already present at \(J_\chi=0\), increasing \(|J_\chi|\) can drive the favored \(\Etwo\) component below an initially leading \(\Aone\) state, thereby changing the symmetry of the lowest \(\Gamma\)-point bound state. More strikingly, for parameters where no \(\Gamma\)-point bound state is present at \(J_\chi=0\), a sufficiently large scalar-chirality coupling can generate one. Scalar chirality therefore does not merely split an existing pair level; it can both select the chirality of the leading \(\Gamma\)-point pair and induce \(\Gamma\)-point two-magnon binding where the Heisenberg interactions alone are insufficient.

This behavior is summarized over the exchange-parameter plane in Fig.~\ref{fig:Jchi_contours}, which shows contours of fixed binding energy \(\Delta_2=10^{-2}\) for several values of \(|J_\chi|\). Choosing a finite binding energy avoids the numerical singularity associated with the exact \(\Delta_2\to0^+\) threshold and also suppresses the strongest finite-size effects. For both signs of \(J_1\), increasing \(|J_\chi|\) substantially enlarges the region supporting a robust \(\Gamma\)-point chiral \(\Etwo\) bound state. Because the one-magnon spectrum is independent of \(J_\chi\), this expansion originates entirely from the modification of the two-magnon interaction.
  
The combined results therefore identify the underlying microscopic mechanism. At \(\Ktot=0\), scalar chirality leaves the \(\Aone\) sector unchanged but shifts the first-neighbor couplings \(V_{1+}\) and \(V_{1-}\) in the two chiral \(\Etwo\) sectors in opposite directions, as seen directly from Eq.~\eqref{eq:V1pm}. The resulting splitting selectively lowers one chiral component without altering the one-magnon continuum. Scalar chirality can consequently enhance the two-magnon binding energy, change the symmetry of the leading \(\Gamma\)-point bound state, and extend the region in which a two-magnon instability can preempt conventional one-magnon condensation.

\subsection{Implications for spin-nematic order}
\label{sec:nematic_order}

The symmetry of the two-magnon bound states constrains
the nature of a possible spin-nematic condensate near
saturation. For spin-$1/2$ moments, the relevant transverse
quadrupolar order parameter is defined on bonds,
\begin{equation}
  \Phi_{\boldsymbol\delta}(\mathbf r)
  =
  \left\langle
    \hat S_{\mathbf r}^-
    \hat S_{\mathbf r+\boldsymbol\delta}^-
  \right\rangle .
\end{equation}
Under a spin rotation about the field axis, this order
parameter transforms with spin angular momentum two.
Its condensation therefore breaks the residual spin
$U(1)$ symmetry down to $\mathbb Z_2$, while transverse
dipolar order can remain absent.

For a bound state at total momentum $\Ktot=0$, the
$\Aone$ channel corresponds to a translationally
uniform bond-nematic order parameter with equal
amplitudes on symmetry-related bonds.
Condensation in a chiral $\Etwopm$ channel instead
produces a bond-quadrupolar order parameter
transforming as $d_{x^2-y^2}\pm i d_{xy}$,
the two chiral combinations of the $E_2$
representation.
For three bond directions related by successive
$2\pi/6$ rotations, the relative amplitudes are
\begin{equation}
  \left(
    \Phi_{\dd_1^{(n)}},
    \Phi_{\dd_2^{(n)}},
    \Phi_{\dd_3^{(n)}}
  \right)
  \propto
  (1,\omega,\omega^2)
  \quad\text{or}\quad
  (1,\omega^2,\omega),
\end{equation}
where $\omega=e^{2\pi i/3}$.
The two patterns exhibit opposite phase windings
under lattice rotations, corresponding to lattice angular momenta
$m=\pm2$ modulo six. A sixfold lattice rotation
changes the order parameter by a phase that can be
compensated by a spin rotation about the field axis.
Thus, the chiral character is encoded in the relative
phases of the bond quadrupoles rather than necessarily
in an anisotropy of their magnitudes. For finite
$J_\chi$, the preferred handedness is selected by
the scalar-chirality interaction itself.

The finite-momentum bound states identified by exact
diagonalization suggest the possibility of spatially
varying spin-nematic order. However, their detailed
bond structure and the selection between single- and
multiple-wave-vector condensates remain undetermined.
More generally, these symmetry considerations
characterize possible condensates, but do not establish
their thermodynamic stability. This requires accounting
for interactions between bound pairs and competing
higher-magnon instabilities.

\section{Conclusion}
\label{sec:conclusion}

We have studied one- and two-magnon excitations above the fully
polarized state of the spin-\(1/2\) triangular-lattice
\(J_1\)-\(J_2\)-\(J_3\) Heisenberg model in the presence of a uniform
scalar-chirality interaction. Our main result is that scalar chirality
acts very differently in the one- and two-magnon sectors. For the
uniform chirality pattern considered here, its contribution cancels
exactly for a single flipped spin, leaving the one-magnon dispersion
and the corresponding instability field unchanged. In the two-magnon
sector, however, this cancellation is incomplete when the two flipped
spins occupy neighboring sites, and the chirality term generates a
genuine two-particle interaction.

At zero total momentum, this effect is particularly transparent in the
symmetry-adapted partial-wave basis. The fully symmetric \(\Aone\)
sector is unaffected by \(J_\chi\), while scalar chirality splits the
two components of the \(\Etwo\) doublet into opposite chiral sectors,
lowering one component and raising the other as shown in Eq.~\eqref{eq:V1pm}. Since the one-magnon
continuum is independent of \(J_\chi\), the resulting increase of the
binding energy can be attributed directly to the modification of the
two-magnon interaction. Depending on the exchange parameters, this
mechanism can strengthen an existing bound state, change the symmetry
of the leading \(\Gamma\)-point bound state from \(\Aone\) to a chiral
\(\Etwo\) component, or generate a \(\Gamma\)-point bound state where
none is present at \(J_\chi=0\). In this respect, scalar chirality
provides a chiral analogue of the pair-enhancing mechanisms generated
by cyclic exchange in frustrated magnets \cite{shannon2006,momoi2006octupolar}.

The Heisenberg model itself exhibits a strong dependence of the
two-magnon binding on the structure of the one-magnon minima. For
\(J_1=-1\), particularly strong binding occurs at
\((J_1,J_2,J_3)=(-1,1,-1/2)\), where the one-magnon dispersion realizes
the \(\mathcal{M}_{\Btwo}\) nodal-line minimum manifold and the leading
\(\Gamma\)-point bound state belongs to the \(\Etwo\) sector. For
\(J_1=+1\), strong binding is found near \((1,-1,-1)\), where the
leading \(\Gamma\)-point state instead belongs to the fully symmetric
\(\Aone\) sector. These results show that both the strength and the
symmetry of magnon pairing depend sensitively on the geometry of the
low-energy one-magnon manifold.

Momentum-resolved exact diagonalization further shows that the minimum
of the two-magnon bound-state dispersion need not occur at \(\Gamma\).
Levels below the continuum are also found at \(M\) and at
incommensurate total momenta, with the latter occurring mainly where
the one-magnon dispersion has incommensurate minima. The \(\Ktot=0\)
gap equation therefore captures an important but not exhaustive part
of the two-magnon problem. Extending the partial-wave analysis to
finite total momentum would provide a natural next step.

Finally, a two-magnon bound state by itself does not in general establish a spin-nematic instability. Near saturation, the relevant quantity is the energy per flipped spin: a two-magnon state preempts conventional one-magnon condensation only if its energy per flipped spin is lower than that of the lowest one-magnon excitation. 
For a \(\Gamma\)-point state, any finite binding automatically makes the pair energetically favorable over a single magnon on a per-flipped-spin basis. At finite total momentum, by contrast, a state may be bound relative to the continuum at that momentum without lying below twice the minimum one-magnon energy.

Even when the two-magnon instability occurs first, bound states of
three or more magnons may become unstable at still higher fields. This
qualification is especially relevant on the triangular lattice, where
higher multipolar phases are known to occur in related models. Our
results therefore identify scalar chirality as a microscopic mechanism
for selectively strengthening two-magnon pairing in a definite
relative-motion chirality and for promoting quadrupolar instabilities
near saturation -- time-reversal-odd interaction can promote a
time-reversal-even instability. Determining the ultimate ordered phase below
saturation, however, requires extending the few-magnon analysis beyond
the two-particle sector.

\begin{acknowledgments}
This work was supported by the National Research, Development and
Innovation Office of Hungary (NKFIH) through OTKA Grant No.~K~142652
and by the Doctoral Students' Excellence Program DKÖP-26-1-BME-28,
and in part by the National Science Foundation through Grant
No.~PHY-2309135 to the Kavli Institute for Theoretical Physics (KITP)
and by COST Action CA23134 (POLYTOPO).
Part of this work was carried out at the Aspen Center for Physics,
which is supported by a grant from the Simons Foundation
(No.~1161654, Troyer).
We thank A.~L.~Chernyshev and J.~Romhányi for discussions that led to
this work, and N.~Shannon for helpful discussions on spin-nematic
physics.
\end{acknowledgments}

\begin{appendix}

\section{The triangular lattice and the lattice harmonics}
\label{sec:trilatt}

Our model describes spins on the triangular lattice illustrated in Fig.~\ref{fig:lattice}. The primitive vectors are
\begin{equation}
    \mathbf{a}_{1}=(1,0),\qquad
    \mathbf{a}_{2}=\left(\frac{1}{2},\frac{\sqrt{3}}{2}\right),
    \label{eq:lattice_vectors}
\end{equation}
with lattice spacing set to unity. Correspondingly, the primitive reciprocal vectors are
\begin{equation}
\mathbf b_1 =
2\pi\left(1,-\frac{1}{\sqrt3}\right),
\qquad
\mathbf b_2 =
2\pi\left(0,\frac{2}{\sqrt3}\right).
\label{eq:triangular_reciprocal}
\end{equation}

\subsection{Point-group symmetry and two-magnon channels}

The point group of the triangular lattice is \(D_6\) (equivalently,
\(C_{6v}\)), generated by a counterclockwise rotation \(R\) by
\(\pi/3\) and a reflection. We take the \(y\)-axis reflection to
belong to the \(\sigma_v\) class. With this convention, the harmonic
\(\beta^{(1)}_{\qq}\) in Eq.~\eqref{eq:B2_harmonic_def} transforms
as \(\Btwo\). It is odd under \(R\) and under reflection about the
\(y\) axis, and even under reflection about the \(x\) axis.
Its nodal set defines the one-magnon minimum manifold
\(\mathcal{M}_{\Btwo}\), which is invariant under \(D_6\).

Exchange of the two magnons reverses their relative momentum,
so \(\psi_{\Ktot}(\qq)=\psi_{\Ktot}(-\qq)\).
On each of the first three neighbor shells, the three even bond
harmonics decompose as \(\Aone\oplus\Etwo\) under \(D_6\).
The \(\Aone\) harmonic is fully symmetric, while the two components
of \(\Etwo\) transform as a planar quadrupole, with the symmetry of
\(q_x^2-q_y^2\) and \(2q_xq_y\) near \(\Gamma\).
At \(\Ktot=0\) and \(J_\chi=0\), the three neighbor shells give
one \(3\times3\) \(\Aone\) block and two identical \(3\times3\)
blocks for the two components of \(\Etwo\).
At nonzero total momentum, the channels are classified by the
little group of \(\Ktot\).

The scalar-chirality interaction breaks lattice reflections
and preserves rotations, reducing the spatial point group from
\(D_6\) to \(C_6\). Under this reduction, \(\Aone\) restricts to
the trivial representation and \(\Etwo\) splits into two
one-dimensional representations. We label these \(\Etwop\) and
\(\Etwom\) according to the harmonics \(\gamma^{(n+)}\) and
\(\gamma^{(n-)}\) defined in Eq.~\eqref{eq:C6harmonics}.
For the active action \((U_R f)(\qq)=f(R^{-1}\qq)\), these
harmonics obey
\begin{equation}
  (U_R\gamma^{(n\pm)})(\qq)
  = \gamma^{(n\pm)}_{R^{-1}\qq}
  = e^{\mp 2\pi i/3}\gamma^{(n\pm)}_{\qq}.
\end{equation}
Equivalently, evaluating the harmonics at \(R\qq\) gives the
phases \(e^{\pm 2\pi i/3}\).
The \(C_3\) generator \(R^2\) has distinct eigenvalues
\(1\), \(e^{-4\pi i/3}\), and \(e^{+4\pi i/3}\) in the
\(\Aone\), \(\Etwop\), and \(\Etwom\) channels, respectively.
Thus a \(C_3\)-symmetric momentum mesh suffices to keep these
three channels decoupled in the \(\Ktot=0\) gap equation.

\subsection{Periodic clusters and momentum meshes}

We use two families of periodic clusters with \(N=L^2\) and
\(N=3L^2\) sites, where \(L\) is a positive integer. Both supercell
translation lattices are invariant under \(D_6\).

\begin{subequations}
\label{eq:periodic_supercells}
For the \(N=L^2\) clusters, the periodicity vectors are
\begin{equation}
\mathbf A_1^{(L^2)} = L\mathbf a_1,
\qquad
\mathbf A_2^{(L^2)} = L\mathbf a_2.
\end{equation}
For the \(N=3L^2\) clusters, they are
\begin{equation}
\mathbf A_1^{(3L^2)} = L(\mathbf a_1+\mathbf a_2),
\qquad
\mathbf A_2^{(3L^2)} = L(-\mathbf a_1+2\mathbf a_2).
\end{equation}
\end{subequations}

The corresponding reciprocal supercell vectors, defined by
\(\mathbf G_i^{(s)}\cdot\mathbf A_j^{(s)}=2\pi\delta_{ij}\),
are
\begin{align}
\mathbf G_1^{(L^2)} &= \frac{\mathbf b_1}{L},
&
\mathbf G_2^{(L^2)} &= \frac{\mathbf b_2}{L},
\nonumber\\
\mathbf G_1^{(3L^2)} &= \frac{2\mathbf b_1+\mathbf b_2}{3L},
&
\mathbf G_2^{(3L^2)} &= \frac{-\mathbf b_1+\mathbf b_2}{3L},
\label{eq:reciprocal_supercells}
\end{align}
where \(\mathbf b_1\) and \(\mathbf b_2\) are given in
Eq.~\eqref{eq:triangular_reciprocal}.

For either family, the mesh \(\mathcal Q_N^{(\sigma)}\) consists of
the periodic momenta shifted by
\begin{equation}
\boldsymbol\eta_\sigma
=
\frac{\sigma}{3}(2\mathbf G_1+\mathbf G_2),
\qquad \sigma=0,\pm1,
\label{eq:mesh_shift_reciprocal}
\end{equation}
where the family superscript on \(\mathbf G_i\) is suppressed.
These shifts are distinct modulo the reciprocal supercell lattice.
The mesh points can be written as
\begin{align}
\mathbf k_{mn}^{(L^2,\sigma)}
&=
\frac{1}{L}
\left[
\left(m+\frac{2\sigma}{3}\right)\mathbf b_1
+
\left(n+\frac{\sigma}{3}\right)\mathbf b_2
\right],
\label{eq:mesh_L2}
\\
\mathbf k_{mn}^{(3L^2,\sigma)}
&=
\frac{1}{3L}
\left[
(2m-n+\sigma)\mathbf b_1
+
(m+n+\sigma)\mathbf b_2
\right].
\label{eq:mesh_3L2}
\end{align}
For the \(L^2\) mesh, \(m,n=0,\ldots,L-1\). For the \(3L^2\)
mesh, \(m=0,\ldots,3L-1\) and \(n=0,\ldots,L-1\). These ranges
give one representative per momentum modulo primitive reciprocal
lattice vectors.

The \(\sigma=0\) mesh implements periodic Bloch quantization.
For general \(\sigma\), the mesh obeys the twisted Bloch
quantization conditions
\begin{equation}
(\mathbf k\cdot\mathbf A_1,\mathbf k\cdot\mathbf A_2)
=
\left(\frac{4\pi\sigma}{3},\frac{2\pi\sigma}{3}\right)
\pmod{2\pi}.
\label{eq:twisted_mesh_bc}
\end{equation}

Let \(R\) be the counterclockwise rotation by \(\pi/3\) defined
above. For both families, \(R^2\mathbf G_1=\mathbf G_2\) and
\(R^2\mathbf G_2=-\mathbf G_1-\mathbf G_2\), so
\begin{equation}
R^2\boldsymbol\eta_\sigma-\boldsymbol\eta_\sigma
=
-\sigma\mathbf G_1.
\end{equation}
Since the difference is a supercell reciprocal vector, each
mesh is invariant under \(C_3\). For \(\sigma=\pm1\), the shift
lies along a mirror axis of the supercell, at angle \(0\) for
the \(L^2\) family and \(\pi/6\) for the \(3L^2\) family,
measured from the \(x\) axis modulo \(\pi\). This reflection
and \(R^2\) generate the \(D_3\) symmetry of each shifted mesh.
The rotation \(R\) exchanges the two shifted meshes because
\(R\boldsymbol\eta_\sigma-\boldsymbol\eta_{-\sigma}
=\sigma(\mathbf G_1+\mathbf G_2)\). Their union is therefore
invariant under \(D_6\). Inversion also exchanges the two meshes,
so they give identical sums for the inversion-even integrands
in the \(\Ktot=0\) gap equation.

For a given \(L\), the unshifted \(3L^2\) mesh is the disjoint
union of the three \(L^2\) meshes,
\begin{equation}
\mathcal Q_{3L^2}^{(0)}
=
\mathcal Q_{L^2}^{(0)}
\,\dot\cup\,
\mathcal Q_{L^2}^{(+1)}
\,\dot\cup\,
\mathcal Q_{L^2}^{(-1)}.
\label{eq:mesh_relation}
\end{equation}
Indeed, the shift \(\boldsymbol\eta_{+1}\) of the \(L^2\) mesh
equals \(\mathbf G_1^{(3L^2)}\) and generates the three cosets
of the \(L^2\) reciprocal supercell lattice in the \(3L^2\)
reciprocal supercell lattice.

We use the shifted meshes to assess the sensitivity of the
\(\Ktot=0\) gap equation to Brillouin-zone sampling, particularly
near weak-binding thresholds. The twists in
Eq.~\eqref{eq:twisted_mesh_bc} specify only the sampling points
in these sums. All exact-diagonalization calculations use
periodic boundary conditions.

\subsection{Lattice harmonics}

We define the first-neighbor bond vectors as
\begin{align}
 \dd_{1}^{(1)} &= -\dd_{4}^{(1)} = \mathbf{a}_{1}, \nonumber\\
 \dd_{2}^{(1)} &= -\dd_{5}^{(1)} = \mathbf{a}_{2}, \nonumber\\
 \dd_{3}^{(1)} &= -\dd_{6}^{(1)} = -\mathbf{a}_{1} + \mathbf{a}_{2}, 
 \label{eq:dd1}
\end{align}
second-neighbor bond vectors as
\begin{align}
 \dd_{1}^{(2)} &= -\dd_{4}^{(2)} =  -\mathbf{a}_{1} + 2 \mathbf{a}_{2} , \nonumber\\
 \dd_{2}^{(2)} &= -\dd_{5}^{(2)} = -2 \mathbf{a}_{1} + \mathbf{a}_{2}, \nonumber\\
 \dd_{3}^{(2)} &= -\dd_{6}^{(2)} = - \mathbf{a}_{1} - \mathbf{a}_{2}, 
 \label{eq:dd2}
\end{align}
and the third-neighbor bond vectors as
\begin{equation}
   \dd_{j}^{(3)} = 2 \dd_{j}^{(1)} .
   \label{eq:dd3}
\end{equation}

Because the two-magnon relative wave function is even under
\(\qq\to-\qq\), only even bond harmonics are required in the present
problem. For each neighbor shell, we define symmetry-adapted lattice
harmonics as Fourier sums over the corresponding bond vectors.
The fully symmetric harmonics are
\begin{equation}
  \gamma^{(n)}_{\qq} = \frac{1}{6} \sum_{j=1}^6 e^{i \qq \cdot \dd_{j}^{(n)} }
\end{equation}
which give
\begin{subequations}
\label{eq:lattice_harmonics_A1}
\begin{align}
\gamma^{(1)}_{\qq}&=\frac{1}{3} \cos q_{x} + \frac{2}{3} \cos\frac{q_{x}}{2}\cos\frac{\sqrt{3}q_{y}}{2},
\\
\gamma^{(2)}_{\qq}&=\frac{1}{3} \cos \sqrt{3}q_{y} + \frac{2}{3} \cos\frac{3q_{x}}{2}\cos\frac{\sqrt{3}q_{y}}{2},
\\
\gamma^{(3)}_{\qq}&=\gamma^{(1)}_{2\qq}.
\label{eq:A1_harmonics}
\end{align}
\end{subequations}
The three even bond harmonics on each neighbor shell decompose under the parent \(D_6\) point group as \(\Aone\oplus\Etwo\). In addition to the fully symmetric harmonics above, we therefore introduce the following symmetry-adapted \(\Etwo\) cosine doublet:
\begin{subequations}
\begin{align}
  \gamma^{(na)}_{\qq}&= \frac{1}{3} 
  \left( 
     2 \cos \qq \cdot \dd_{1}^{(n)} 
     - \cos \qq \cdot \dd_{2}^{(n)}
     - \cos \qq \cdot \dd_{3}^{(n)}
      \right) \\
  \gamma^{(nb)}_{\qq}&= \frac{1}{\sqrt{3}} 
    \left( 
       \cos \qq \cdot \dd_{2}^{(n)}
     - \cos \qq \cdot \dd_{3}^{(n)}
      \right) 
\end{align}
\end{subequations}
Explicitly,
\begin{align}
\gamma^{(1a)}_{\qq}&=\frac{2}{3}\left(\cos q_{x}-\cos\frac{q_{x}}{2}\cos\frac{\sqrt{3}q_{y}}{2}\right),
\nonumber\\
\gamma^{(2a)}_{\qq}&=\frac{2}{3}\left(\cos \sqrt{3}q_{y} -\cos\frac{3q_{x}}{2}\cos\frac{\sqrt{3}q_{y}}{2}\right),
\nonumber\\
\gamma^{(3a)}_{\qq}&=\gamma^{(1a)}_{2\qq},
\nonumber\\
\gamma^{(1b)}_{\qq}&=-\frac{2}{\sqrt{3}}\sin\frac{q_{x}}{2}\sin\frac{\sqrt{3}q_{y}}{2},
\nonumber\\
\gamma^{(2b)}_{\qq}&=\frac{2}{\sqrt{3}}\sin\frac{3q_{x}}{2}\sin\frac{\sqrt{3}q_{y}}{2},
\nonumber\\
\gamma^{(3b)}_{\qq}&=\gamma^{(1b)}_{2\qq},
\label{eq:other_harmonics}
\end{align}
Here, \(\gamma^{(na)}_{\qq}\) and \(\gamma^{(nb)}_{\qq}\) denote the two components of the \(\Etwo\) doublet in the \(n\)th neighbor shell. These symmetry-adapted lattice harmonics satisfy the decomposition identity
\begin{equation}
  \gamma^{(n)}_{\qq+\pp}
  +
  \gamma^{(n)}_{\qq-\pp}
  =
  2\gamma^{(n)}_{\qq}\gamma^{(n)}_{\pp}
  +
  \gamma^{(na)}_{\qq}\gamma^{(na)}_{\pp}
  +
  \gamma^{(nb)}_{\qq}\gamma^{(nb)}_{\pp}.
  \label{eq:decompose}
\end{equation}

For channel decoupling in the chiral phase, it is convenient to
introduce the chiral \(\Etwo\) combinations
\begin{equation}
\gamma^{(n\pm)}_{\qq}
=
\gamma^{(na)}_{\qq}
\pm i\gamma^{(nb)}_{\qq}.
\label{eq:C6harmonics}
\end{equation}
The two functions are complex conjugates,
\(\gamma_{\qq}^{(n-)}=[\gamma_{\qq}^{(n+)}]^*\), and form
one-dimensional representations of \(C_6\) carrying opposite
lattice angular momenta \(m=\pm2\) modulo six.
Thus, upon reducing the point-group symmetry from \(D_6\) to \(C_6\),
the \(\Etwo\) doublet decomposes into these two one-dimensional
components.
Equivalently,
\begin{align}
  \gamma^{(n\pm)}_{\qq}
  &=
  \frac{1}{3}
  \sum_{j=1}^{6}
  e^{\pm 2 i (j-1)\pi/3}
  e^{i\qq\cdot\dd_{j}^{(n)}}
  \nonumber\\
  &=
  \gamma^{(na)}_{\qq}
  \pm i\gamma^{(nb)}_{\qq}.
\end{align}
The decomposition rule is now
\begin{equation}
  \gamma^{(n)}_{\qq+\pp}
  +
  \gamma^{(n)}_{\qq-\pp}
  =
  2\gamma^{(n)}_{\qq}\gamma^{(n)}_{\pp}
  + \frac{1}{2}
  \bar \gamma^{(n-)}_{\qq}\gamma^{(n-)}_{\pp}
  + \frac{1}{2}
  \bar \gamma^{(n+)}_{\qq}\gamma^{(n+)}_{\pp}.
   \label{eq:decomposeChi}
 \end{equation}

\section{Auxiliary-space derivation of the chiral two-magnon kernel}
\label{app:chirality_kernel}

This appendix presents the technical derivation of the scalar-chirality contribution to the two-magnon interaction kernel. The purpose of the derivation is to isolate the genuine two-magnon interaction from the free propagation of two independent spin flips while keeping the spin-\(1/2\) hard-core constraint explicit.

We work throughout with excitation energies measured relative to the fully polarized state. Let \(\mathcal{H}_{2\mathrm R}\) denote the physical two-magnon Hilbert space,
\begin{equation}
    \mathcal{H}_{2\mathrm R}
    =
    \{\ket{i,j}\,|\,i\neq j\},
    \qquad
    \ket{i,j}
    =
    \hat S_j^-\hat S_i^-\ket{\mathrm{FM}},
\end{equation}
and enlarge it by adjoining auxiliary double-occupancy states,
\begin{equation}
    \mathcal{H}_{2\mathrm D}
    =
    \{\ket{j,j}\,|\,j=1,\ldots,N\}.
\end{equation}
The enlarged two-particle space is
\begin{equation}
    \mathcal{H}_{2\mathrm X}
    =
    \mathcal{H}_{2\mathrm R}
    \oplus
    \mathcal{H}_{2\mathrm D}.
    \label{eq:app_extended_space}
\end{equation}
The states in \(\mathcal{H}_{2\mathrm D}\) are not physical spin-\(1/2\) states; they are introduced only as auxiliary bookkeeping states that allow the independent propagation of the two spin-flip coordinates to be defined before the hard-core projection is imposed. We refer to \(\mathcal{H}_{2\mathrm R}\) as the regular (physical) sector and \(\mathcal{H}_{2\mathrm D}\) as the double-occupancy sector.

At fixed total momentum \(\Ktot\), we use the non-normalized regular state
\begin{equation}
    \ket{\Ktot,\qq}_{\mathrm R}
    =
    \sum_{i\neq j}
    e^{-i\frac{\Ktot}{2}\cdot(\mathbf r_i+\mathbf r_j)}
    e^{-i\qq\cdot(\mathbf r_j-\mathbf r_i)}
    \ket{i,j},
    \label{eq:app_regular_state}
\end{equation}
together with the auxiliary double-occupancy state
\begin{equation}
    \ket{\Ktot}_{\mathrm D}
    =
    \sum_{j=1}^N
    e^{-i\Ktot\cdot\mathbf r_j}
    \ket{j,j}.
    \label{eq:app_double_state}
\end{equation}
Their sum,
\begin{equation}
    \ket{\Ktot,\qq}_{\mathrm X}
    =
    \ket{\Ktot,\qq}_{\mathrm R}
    +
    \ket{\Ktot}_{\mathrm D}.
    \label{eq:app_extended_state}
\end{equation}
is the corresponding unrestricted two-particle momentum state. The projector onto the regular two-magnon sector is therefore
\begin{equation}
    \hat{\mathcal{P}}_{2\mathrm R}
    =
    \hat{\mathbf 1}_{2\mathrm X}
    -
    \sum_j
    \ket{j,j}\bra{j,j}.
    \label{eq:app_regular_projector}
\end{equation}

The role of the double-occupancy sector is to separate the free two-magnon propagation from the genuine interaction terms. We define the free two-magnon Hamiltonian \(\hat H_0\) in the enlarged space by its action on the momentum states,
\begin{equation}
    \hat H_0\ket{\Ktot,\qq}_{\mathrm X}
    =
    \left(
    \varepsilon_{\frac{\Ktot}{2}+\qq}
    +
    \varepsilon_{\frac{\Ktot}{2}-\qq}
    +
    2E_{0}
    \right)
    \ket{\Ktot,\qq}_{\mathrm X},
    \label{eq:app_H0_def}
\end{equation}
which can be realized for any arbitrary \(\hat{H}\) by
\begin{equation}
    \hat H_0 \left( \ket{i}_{1} \otimes \ket{j}_{2} \right)
    =
    \left( \hat{H}\ket{i}_{1} \right) \otimes \ket{j}_{2} + \ket{i}_{1} \otimes \left( \hat{H}\ket{j}_{2} \right),
    \label{eq:app_H0_rel}
\end{equation}
where \(\ket{i}_{1} \otimes \ket{j}_{2} = \ket{i,j}\). Therefore, in real space this operator acts independently on the two spin-flip coordinates before the hard-core projection is imposed. Assuming that the original Hamiltonian can be decomposed into a free part and an interaction part, the latter can be obtained by taking the difference between the full and free Hamiltonians:
\begin{equation}
    \hat V
    =
    \hat H
    -
    \hat H_0.
    \label{eq:app_V_def}
\end{equation}
This interaction operator acts on both the regular and double-occupancy sectors; however, only its projection onto the regular sector contributes to the final interaction kernel. Using Eqs.~\eqref{eq:app_extended_state} and \eqref{eq:app_V_def}, Eq.~\eqref{eq:app_H0_def} can be rewritten as
\begin{align}
    \left(\hat{H} - \hat{V}\right)
    &\ket{\Ktot,\qq}_{\mathrm R}
    +
    \hat H_0 \ket{\Ktot}_{\mathrm D}
    =\nonumber\\
    &\left(
    \varepsilon_{\frac{\Ktot}{2}+\qq}
    +
    \varepsilon_{\frac{\Ktot}{2}-\qq}
    +
    2E_{0}
    \right)
    \left(
    \ket{\Ktot,\qq}_{\mathrm R}
    +
    \ket{\Ktot}_{\mathrm D}
    \right)
    \label{eq:app_first_substitution}
\end{align}

The action of \(\hat{H}\) on a regular momentum state is given by
\begin{align}
    \hat{H}\ket{\Ktot,\qq}_{\mathrm R}
    =&
    \left[
    \varepsilon_{\frac{\Ktot}{2}+\qq}
    +
    \varepsilon_{\frac{\Ktot}{2}-\qq}
    +
    E_{0}
    \right]
    \ket{\Ktot,\qq}_{\mathrm R}
    +\nonumber\\
    &\frac{1}{N}
    \sum_{\pp}
    V_{\Ktot}(\qq,\pp)^\mathsf{T}
    \ket{\Ktot,\pp}_{\mathrm R}.
    \label{eq:app_regular_action}
\end{align}
Rather than starting from the general two-magnon wave function
in Eq.~\eqref{eq:two_magnon_wavefunction}, we work directly with
the states defined in Eqs.~\eqref{eq:app_regular_state},
\eqref{eq:app_double_state}, and~\eqref{eq:app_extended_state},
which makes the separation of the regular and double-occupancy
contributions more transparent.
In Eq.~\eqref{eq:app_regular_action}, \(\qq\) labels the initial
relative momentum and \(\pp\) the final one, whereas their roles
are reversed in the Schr\"odinger
equation~\eqref{eq:two_magnon_schrodinger}.
This accounts for the appearance of the transposed interaction
kernel \(V_{\Ktot}(\qq,\pp)^\mathsf{T}=V_{\Ktot}(\pp,\qq)\).

Applying Eq.~\eqref{eq:app_regular_action} to Eq.~\eqref{eq:app_first_substitution}, we obtain the following expression for the effective potential associated with the Schr\"odinger equation:
\begin{align}
    \frac{1}{N}
    \sum_{\pp}
    V_{\Ktot}(\qq,\pp)&
    ^\mathsf{T}\ket{\Ktot,\pp}_{\mathrm R}
    =\nonumber\\
    &\left( \hat{V} + E_{0} \right)
    \ket{\Ktot,\qq}_{\mathrm R}
    -
    \hat H_0
    \ket{\Ktot}_{\mathrm D}
    +
    \nonumber\\
    &\left(
    \varepsilon_{\frac{\Ktot}{2}+\qq}
    +
    \varepsilon_{\frac{\Ktot}{2}-\qq}
    +
    2E_{0}
    \right)
    \ket{\Ktot}_{\mathrm D}
    \label{eq:app_unprojected_final}
\end{align}
Finally, projecting Eq.~\eqref{eq:app_unprojected_final} onto the regular sector, we obtain the final expression used to determine the effective potential:
\begin{align}
    \frac{1}{N}
    \sum_{\pp}
    V_{\Ktot}(\qq,\pp)&
    ^\mathsf{T}\ket{\Ktot,\pp}_{\mathrm R}
    =\nonumber\\
    &\left( \hat{\mathcal{P}}_{2\mathrm R}\hat{V} + E_{0} \right)
    \ket{\Ktot,\qq}_{\mathrm R}
    -
    \hat{\mathcal{P}}_{2\mathrm R}
    \hat H_0
    \ket{\Ktot}_{\mathrm D}
    \label{eq:app_projected_final}
\end{align}
Acting with \(\hat{\mathcal{P}}_{2\mathrm R}(\hat{H}-\hat H_0)+E_{0}\) on the real-space compontents of \(\ket{\Ktot,\qq}_{\mathrm R}\), and Fourier transforming back to the relative momentum \(\pp\), we obtain the contribution of the regular part of the equation. The remaining term, \(\hat{\mathcal{P}}_{2\mathrm R}\hat H_0\ket{\Ktot}_{\mathrm D}\), can be rearranged as
\begin{align}
    \hat{\mathcal{P}}_{2\mathrm R}
    \hat H_0
    \ket{\Ktot}_{\mathrm D}&
    =
    \hat{\mathcal{P}}_{2\mathrm R}
    \hat H_0
    \frac{1}{N}
    \sum_{\pp}
    \ket{\Ktot,\pp}_{\mathrm X}
    =\nonumber\\
    &\hat{\mathcal{P}}_{2\mathrm R}
    \frac{1}{N}
    \sum_{\pp}
    \left(
    \varepsilon_{\frac{\Ktot}{2}+\pp}
    +
    \varepsilon_{\frac{\Ktot}{2}-\pp}
    +
    2E_{0}
    \right)
    \ket{\Ktot,\pp}_{\mathrm X}
    =\nonumber\\
    &\frac{1}{N}
    \sum_{\pp}
    \left(
    \varepsilon_{\frac{\Ktot}{2}+\pp}
    +
    \varepsilon_{\frac{\Ktot}{2}-\pp}
    +
    2E_{0}
    \right)
    \ket{\Ktot,\pp}_{\mathrm R}
    \label{eq:doubly_occupancy_in_final}
\end{align}

For the scalar-chirality term, the projected interaction can be evaluated directly in the configuration basis. As a consequence of the cancelation in the one-magnon sector, the general expression~\eqref{eq:app_projected_final} simplifies considerably for the scalar-chirality term. Using
\begin{equation}
    4\hat\chi_{ijk}
    =
    i\hat P_{ijk}
    -
    i\hat P_{ikj},
    \label{eq:app_chirality_permutation}
\end{equation}
each oriented triangle contributes an imaginary amplitude whose sign is fixed by the orientation of the cyclic permutation. The terms that cancel in the one-magnon sector no longer cancel pairwise in the two-magnon sector.

Substituting Eq.~\eqref{eq:doubly_occupancy_in_final} into Eq.~\eqref{eq:app_projected_final} for scalar-chirality, we obtain
\begin{align}
V_{\Ktot}^{(\chi)}(\qq,\pp)^\mathsf{T}
&=
i\,12\sqrt{3}J_{\chi}
\bigg[
\gamma_{\pp}^{(1)}
\left(
\gamma_{\frac{\Ktot}{2}}^{(1a)}
\gamma_{\qq}^{(1b)}
-
\gamma_{\frac{\Ktot}{2}}^{(1b)}
\gamma_{\qq}^{(1a)}
\right)
\nonumber\\
&\hspace{2.0cm}
+
\gamma_{\pp}^{(1a)}
\left(
\gamma_{\frac{\Ktot}{2}}^{(1b)}
\gamma_{\qq}^{(1)}
-
\gamma_{\frac{\Ktot}{2}}^{(1)}
\gamma_{\qq}^{(1b)}
\right)
\nonumber\\
&\hspace{2.0cm}
+
\gamma_{\pp}^{(1b)}
\left(
\gamma_{\frac{\Ktot}{2}}^{(1)}
\gamma_{\qq}^{(1a)}
-
\gamma_{\frac{\Ktot}{2}}^{(1a)}
\gamma_{\qq}^{(1)}
\right)
\bigg].
\label{eq:app_int_chi_real_harmonics}
\end{align}
Here \(\gamma^{(1)}\), \(\gamma^{(1a)}\), and \(\gamma^{(1b)}\) are the first-neighbor triangular-lattice harmonics defined in the main text or in Appendix~\ref{sec:trilatt}.
The antisymmetric structure in the brackets reflects the orientation of the scalar-chirality interaction.

For the partial-wave decomposition, it is useful to rewrite Eq.~\eqref{eq:app_int_chi_real_harmonics} in the complex \(C_6\) harmonic basis. Using the relations between the real and complex first-neighbor harmonics, one obtains Eq.~\eqref{eq:app_int_chi_complex_harmonics} in the main text, where the transposition is explicitly carried out.

\end{appendix}

\bibliography{ref}

\end{document}